\documentclass[twocolumn]{aastex63}

\usepackage{graphicx}   
\usepackage{subfigure}
\usepackage{multirow}
\usepackage{amsmath}

\graphicspath{{./}{./figures/}}
\shorttitle{impact of cosmic filaments}
\shortauthors{Zhu, Zhang \& Feng.}
\begin{document}

\title{Impact of cosmic filaments on the gas accretion rate of dark matter halos}

\correspondingauthor{Weishan Zhu}
\email{zhuwshan5@mail.sysu.edu.cn}

\author[0000-0002-1189-2855]{Weishan, Zhu}
\affil{School of Physics and Astronomy, Sun Yat-Sen University, Zhuhai campus, No. 2, Daxue Road \\
Zhuhai, Guangdong, 519082, China}
\affil{CSST Science Center for the Guangdong-Hong Kong-Macau Greater Bay Area, Zhuhai, 519082, China}

\author{Fupeng, Zhang}
\affil{School of Physics and Materials Science, Guangzhou University\\
Guangzhou, Guangdong, 510006, China}

\author{Long-Long, Feng}
\affil{School of Physics and Astronomy, Sun Yat-Sen University, Zhuhai campus, No. 2, Daxue Road \\
Zhuhai, Guangdong, 519082, China}
\affil{CSST Science Center for the Guangdong-Hong Kong-Macau Greater Bay Area, Zhuhai, 519082, China}



\begin{abstract}
We investigate the impact of cosmic filaments on the gas accretion rate, $\dot{M}_{\rm{gas}}$, of dark matter halos in filaments, based on cosmological hydrodynamic simulation. We find that for halos less massive than $10^{12.0}\  \rm{M_{\odot}}$, $\dot{M}_{\rm{gas}}$ of halos residing in prominent filaments (with width $D_{\rm{fil}}>3\ \rm{Mpc}/h$) is lower than halos residing in tenuous filaments ($D_{\rm{fil}}<3\ \rm{Mpc}/h$) by $20-30\%$ at $z=0.5$, and by a factor of 2-3 at $z=0$. However, $\dot{M}_{\rm{gas}}$ depends weakly on the physical distance between halo center and the spine of filaments from high redshift to $z=0$ and only shows a clear difference between the inner and outer regions in prominent filaments at $z=0$. We further probe the thermal properties of gas in prominent and tenuous filaments, which appear in relatively highly and intermediate overdense regions, respectively. The gas in prominent filaments is hotter. Around $26\%$, $38\%$ and $45\%$ of gases in prominent filaments are hotter than $10^6$ K at $z=1.0, 0.5$ and $z=0.0$ respectively. The corresponding fractions in tenuous filaments are merely $\sim 6\%, 9\%$ and $11\%$. The suppressed gas accretion rate for low mass halos in prominent filaments at $z \lesssim 0.5$ may result from the hotter ambient gas, which could provide a physical processing mechanism to cut down the supply of gas to halos before they enter clusters. This process meets partially the need of the preheating mechanism implemented in some semi-analytical models of galaxy formation, but works only for $\sim 20\%$ of halos at $z < 1$.

\end{abstract}

\keywords{Large-scale structures --- 
cosmic web --- dark matter halo --- Hydrodynamical simulations}


\section{Introduction} \label{sec:intro}

%\begin{itemize}
%  \item{v6.3}
%   \begin{enumerate}
%      \item New {\tt\string interactive} environment to highlight interactive figures (see Section \ref{animation}),
%      \item Improved collaboration commands, 
%      \item New {\tt\string anonymous} style to keep the authors, affiliations and acknowledgments from showing in the compiled pdf for dual anonymous review, and
%      \item Adoptions of IAU approved syntax for nominal units, see Section \ref{nominal}.
%    \end{enumerate}
%\end{itemize}
Since the pioneering works in 1970s (e.g. \citealt{1970A&A.....5...84Z}; \citealt{1973A&A....27....1I}; \citealt{1979ApJ...231....1W}), theoretical studies on the anisotropic gravitational collapse of cosmic matter have predicted the formation of a web-like appearance of matter distribution on large scales, consisting of structures such as nodes/clusters, filaments, sheets/walls and voids (e.g. \citealt{1996Natur.380..603B}; \citealt{2008LNP...740..335V}). Observational surveys in the past decades have confirmed that the distribution of galaxies at low and intermediate redshifts shows a web-like pattern, in agreement with the theoretical prediction (e.g. \citealt{1986ApJ...302L...1D}; \citealt{2003astro.ph..6581C}; \citealt{2004ApJ...606..702T}; \citealt{2014MNRAS.438..177A}). The evolution of cosmic web and associated properties have also been illustrated and explored with large cosmological simulations, providing more vivid pictures and quantitative insights to the cosmic web in detail (e.g., \citealt{2005Natur.435..629S}; \citealt{2005MNRAS.359..272C}; \citealt{2007A&A...474..315A}: \citealt{2010MNRAS.408.2163A}; \citealt{2014MNRAS.441.2923C}; \citealt{2014MNRAS.444.1453D}; \citealt{2014MNRAS.444.1518V}; \citealt{2015MNRAS.446..521S}; \citealt{2021ApJ...906...68L}; \citealt{2021MNRAS.502..714R}; \citealt{2021PhRvD.103f3517H}).
An important related issue is the impact of large scale cosmic web on the properties of galaxies.

Many observational studies have concluded that the properties of nearby galaxies, such as star formation activity, stellar mass, colour, and morphology, are related to their environment. Early work found that the elliptical and S0 population increases with the local galaxy density, yet the spirals shows a reverse trend (e.g. \citealt{1980ApJ...236..351D}). As density increases, the star formation activity of nearby galaxies decreases significantly  (e.g., \citealt{2004MNRAS.353..713K}; \citealt{2005ApJ...621..201C}; \citealt{2010ApJ...721..193P}; \citealt{2014MNRAS.438..177A}; \citealt{2014MNRAS.439.3564C}). Early-type galaxies, which are redder and more luminous, are found largely in dense regions (\citealt{2003ApJ...585L...5H}; \citealt{2005ApJ...629..143B}; \citealt{2009MNRAS.399..966S}).
In recent years, some observations have also implied that the relation between galaxies properties and environment at low redshifts can be extended to intermediate and high redshifts (e.g. \citealt{2007ApJS..172..284C}; \citealt{2009ApJ...705L..67P}; \citealt{2010ApJ...721..193P};
\citealt{2016ApJ...825...72A};
\citealt{2017ApJ...841L..22G}; \citealt{2017ApJ...847..134K}; \citealt{2020ApJ...890....7C}; \citealt{2020MNRAS.493.5987O}), although some other works report a reverse relation (e.g. \citealt{2007A&A...468...33E}; \citealt{2008MNRAS.383.1058C}), or lack of correlation at intermediate and high redshifts (e.g. \citealt{2008ApJ...684..888P}; \citealt{2011MNRAS.411..929G}; \citealt{2011MNRAS.418..938G}; \citealt{2016ApJ...825..113D}). 

It is now well accepted that the local environments on scale of $\sim 1$ Mpc/h have played important roles in shaping the galaxies properties (e.g. \citealt{2004MNRAS.353..713K}; \citealt{2012MNRAS.420.1481V}), which are primarily determined by the masses of host dark matter halos. In more massive halos, mechanisms such as ram pressure stripping, harassment and starvation/strangulation (e.g. \citealt{1972ApJ...176....1G};
\citealt{1978MNRAS.183..341W};
\citealt{1980ApJ...237..692L};
\citealt{1996Natur.379..613M};
\citealt{1999MNRAS.308..947A};
\citealt{2000ApJ...540..113B}; \citealt{2000Sci...288.1617Q}; \citealt{2009ApJ...694..789T}; \citealt{2015Natur.521..192P};
\citealt{2019MNRAS.489.5582S};
\citealt{2020MNRAS.494.1114S}) are assumed to be more effective to deplete the gas reservoir of galaxies, especially satellite galaxies. Cosmological hydrodynamical simulations have shown that the accretion rate of galaxies, primary satellite, is suppressed significantly in massive halos (e.g. \citealt{2009MNRAS.399..650S}; \citealt{2017MNRAS.466.3460V}).

However, the impact of global environment above 1 Mpc/h and even larger scales on galaxies' properties remains a little bit controversial in the literature. Some works suggested that galaxies' properties, such as the stellar mass function, show no/weak dependence on the global environment (\citealt{2013MNRAS.432.3141C}; \citealt{2013A&A...550A..58V}; \citealt{2018MNRAS.481.3456C}). On the other hand, many recent observational studies show that more massive and passive galaxies tend to reside closer to the large scale filaments. With respect to galaxies in fields and voids, the gas content of galaxies will decrease, and the color of star forming galaxies at fixed mass is becoming more red while approaching to filaments and clusters (e.g. \citealt{2016MNRAS.457.2287A}; \citealt{2017MNRAS.466.1880C}; \citealt{2017ApJ...837...16D}; \citealt{2017A&A...600L...6K};
\citealt{2018MNRAS.478.4336M};
\citealt{2018MNRAS.474..547K};
\citealt{2019A&A...632A..49S};
\citealt{2020A&A...638A..75B};
\citealt{2020MNRAS.497..466S};
\citealt{2021arXiv210513368W}; \citealt{2021arXiv210104389C}). These effects are mainly found at low redshifts, while some works also report signals at $z \sim 0.7-1.0$. Meanwhile, a similar phenomenon has also been found in cosmological hydrodynamical simulations such as EAGLE and Horizon-AGN (e.g. \citealt{2020MNRAS.497.2265S}; \citealt{2020MNRAS.498.1839X}; \citealt{2021MNRAS.501.4635S} ). These studies suggest that the large scale cosmic web could also have effects on the galaxies properties, and galaxies may have undergone pre-processing in cosmic filaments before entering clusters.

The reported transition in properties of galaxies while approaching to cosmic filaments and clusters maybe caused by several mechanisms such as enhanced galaxy merge rate, gas removal via ram pressure stripping, and/or starvation by cut-off of external gas supply (e.g., \citealt{2017A&A...600L...6K}; \citealt{2019OJAp....2E...7A}; \citealt{2021MNRAS.501.4635S}; \citealt{2021arXiv210513368W}). Based on Illustris and IllustrisTNG simulations, \cite{2016MNRAS.457.3024H} and \cite{2019MNRAS.486.3766M} have shown that the intergalactic mediums residing in filaments and nodes are more dense and hotter than those in voids and walls. The different properties of the IGM may lead to difference on the gas supply, and gas stripping of halos and their central galaxies. 

In \citealt{2021ApJ...920....2Z}, we have quantitatively evaluated the evolution and distribution of the local diameter of cosmic filaments in a cosmological hydrodynamical simulation, confirming the rapid growth of prominent filaments after z=2. Moreover, we have probed the density and temperature profile of filaments. The density profiles can be described by an isothermal single-beta model. We found that the typical temperature of baryonic gas in filaments is related with the width of filaments. Gas in thick filaments are hotter than in thin filaments. Therefore, the gas supply of dark matter halos in filaments may be dependent on the width of filaments. In this work, we use samples from a cosmological hydrodynamical simulation to study the impact of cosmic filaments on the gas supply of halos, i.e. the gas accretion rate. 

This work is laid out as follows: The cosmological hydrodynamical simulation used here is briefly introduced in Section 2. The method used to identify filaments and halos, and the procedure used to estimate the properties of filaments and gas accretion rate to halos are also presented in Section 2. Section 3 investigates the dependence of gas accretion rate on the distance between halo center and the spine of filaments, and on the width of filaments. In Section 4, we explore the properties of gas in filaments, aiming to find the probable reason that leads to the dependence of gas accretion rate onto halos on the width of host filaments. We summary our findings and discuss the connection to observations and previous studies in Section 5.

\section{Methodology} \label{sec:method}

\subsection{Simulation Samples} %\label{sec:simulation}
We carry out our investigation using a cosmological hydrodynamical simulation run by the adaptive mesh refinement(AMR) code RAMSES (\citealt{2002A&A...385..337T}). Assuming a $\Lambda$CDM cosmology, this simulation tracks the evolution in a volume of $(100 h^{-1})^3$ Mpc from $z=99$ to $z=0$, with the cosmological parameters $\Omega_{m}=0.317, \Omega_{\Lambda}=0.683,h=0.671,\sigma_{8}=0.834, \Omega_{b}=0.049$, and $n_{s}=0.962$ (\citealt{2014A&A...571A..16P}). The simulation uses $1024^3$ dark matter particles and a $1024^3$ root grid, resulting in mass resolution of $1.03 \times \rm{10^{8}\ M_{\odot}}$, and  spatial resolution of $97.6\ h^{-1}$ kpc for the root grid. The highest AMR grid level is set to $l_{max}=17$, which reaches $0.763\ h^{-1}$ kpc for a grid cell at the finest level. A uniform UV background following the model in \cite{1996ApJ...461...20H} is switched on at $z=8.5$. Radiative cooling and heating of gas, star formation and stellar feedback are included in this simulation, while feedback from active galactic nuclei(AGN) is not included. For more details about the simulation, we referrer the readers to \cite{2021ApJ...906...95Z}.

\subsection{cosmic web classification}

We construct the density of baryonic and dark matter on a $512^3$ grid respectively, based on the simulation samples. Then the grid cells are classified into four categories of cosmic large scale environment, i.e., nodes/clusters, filaments, sheets and voids, using the tidal tensor, i.e., the Hessian matrix, of the rescaled peculiar gravitational potential $\phi$, 

\begin{equation}
   T_{\alpha \beta}=\frac{\partial^{2} \phi}{\partial r_{\alpha} \partial r_{\beta}},
\end{equation}

where $\nabla^2\phi=\delta$, $\delta$ is the density contrast, and $\alpha, \beta=$ 1, 2, 3 indicate the components of coordinate axes. More specifically, the environment of each grid cell is determined by the number of eigenvalues of the Hessian matrix larger than a given threshold value, $\lambda_{\rm{th}}$. Namely, if a cell have 0, 1, 2, or 3 eigenvalues greater than $\lambda_{\rm{th}}$, it will be marked, in turn, as a cell in void, sheet, filament or cluster. Note that a precise theoretical determination of $\lambda_{\rm{th}}$ from the anisotropic collapse of structures is not available in the literature. Recent works often use values of 0.2-0.4, following the suggestion in \cite{2009MNRAS.396.1815F}. In addition, a constant $\lambda_{\rm{th}}$ at different redshifts is usually adopted, which has been shown to be able to capture the structure of cosmic web at different epochs (e.g. \citealt{2017ApJ...838...21Z}; \citealt{2019MNRAS.486.3766M}). In this work, we carry out investigations following our choice in \cite{2021ApJ...920....2Z}, i.e., $\lambda_{\rm{th}}=0.2$ at different redshifts. More details about this web classification scheme and the choice of $\lambda_{\rm{th}}$ can be found in  \cite{2007MNRAS.375..489H}, \cite{2009MNRAS.396.1815F} and \cite{2017ApJ...838...21Z}.

Following the procedures in \cite{2014MNRAS.441.2923C}, the filaments are compressed to find their spine, and then their local widths, $D_{\rm{fil}}$, are estimated. The properties of filament such as the distribution and evolution of local diameter and linear density, as well as the matter density and temperature profiles in filaments have been studied in \cite{2021ApJ...920....2Z}. Since the cosmic web environment is identified by density field of spatial resolution $\sim 200\ h^{-1} \rm{kpc}$, the results presented in work are mainly applicable for filaments with diameters $D_{\rm{fil}} \geq 200\ h^{-1} \rm{kpc}$. Moreover, \cite{2021ApJ...920....2Z} show that, $\sim 95\%$ of the gas residing in filaments are hosted by filaments with $D_{\rm{fil}} \geq 500\ h^{-1} \rm{kpc}$.

\subsection{halos and their environment}
\begin{figure}[htbp]
\begin{center}
%\epsscale{1.5}
%\vspace{-0.5cm}
\includegraphics[width=0.50\textwidth, trim=0 10 10 10, clip]{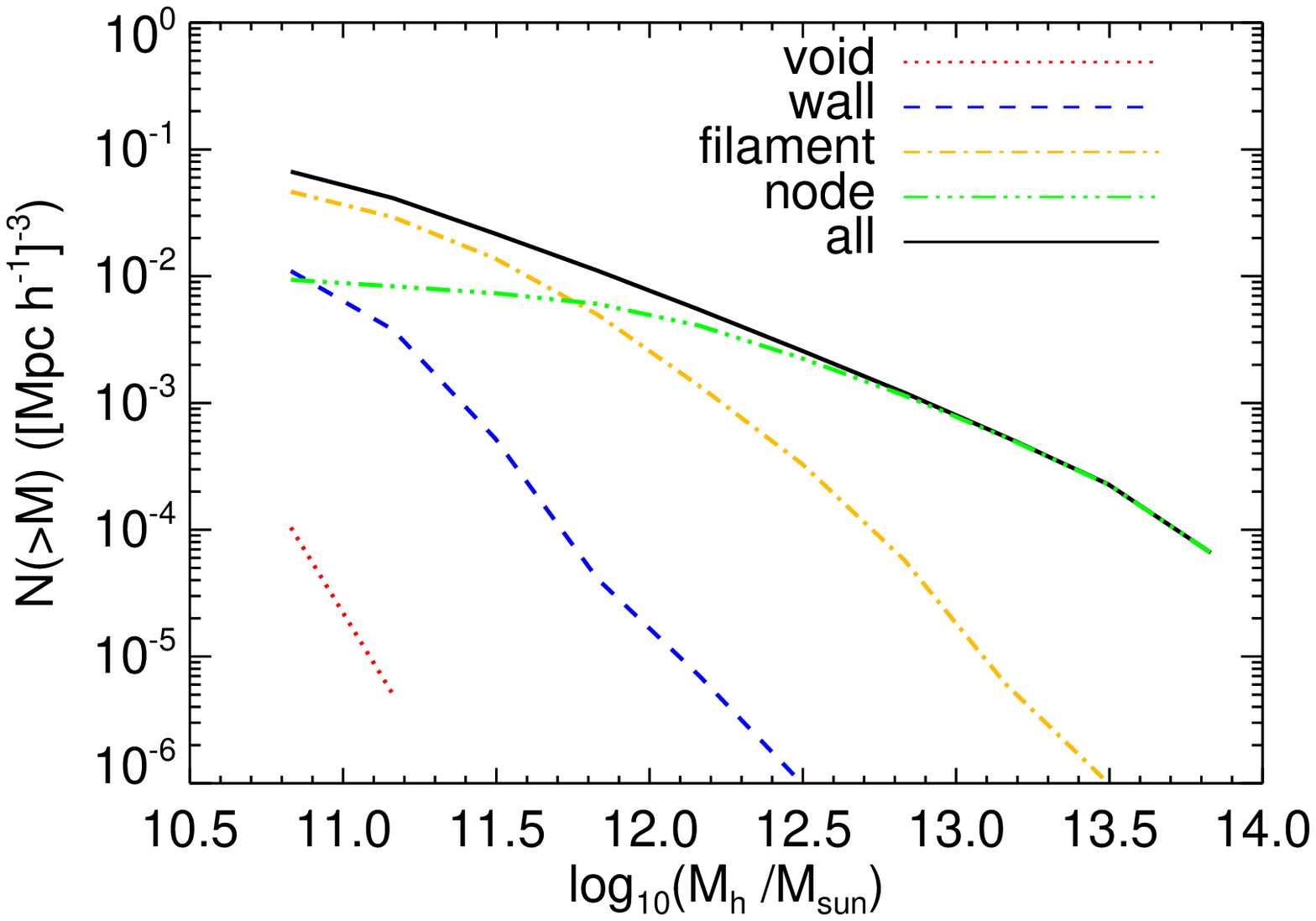}
\includegraphics[width=0.40\textwidth, trim=0 10 10 10, clip, angle=90]{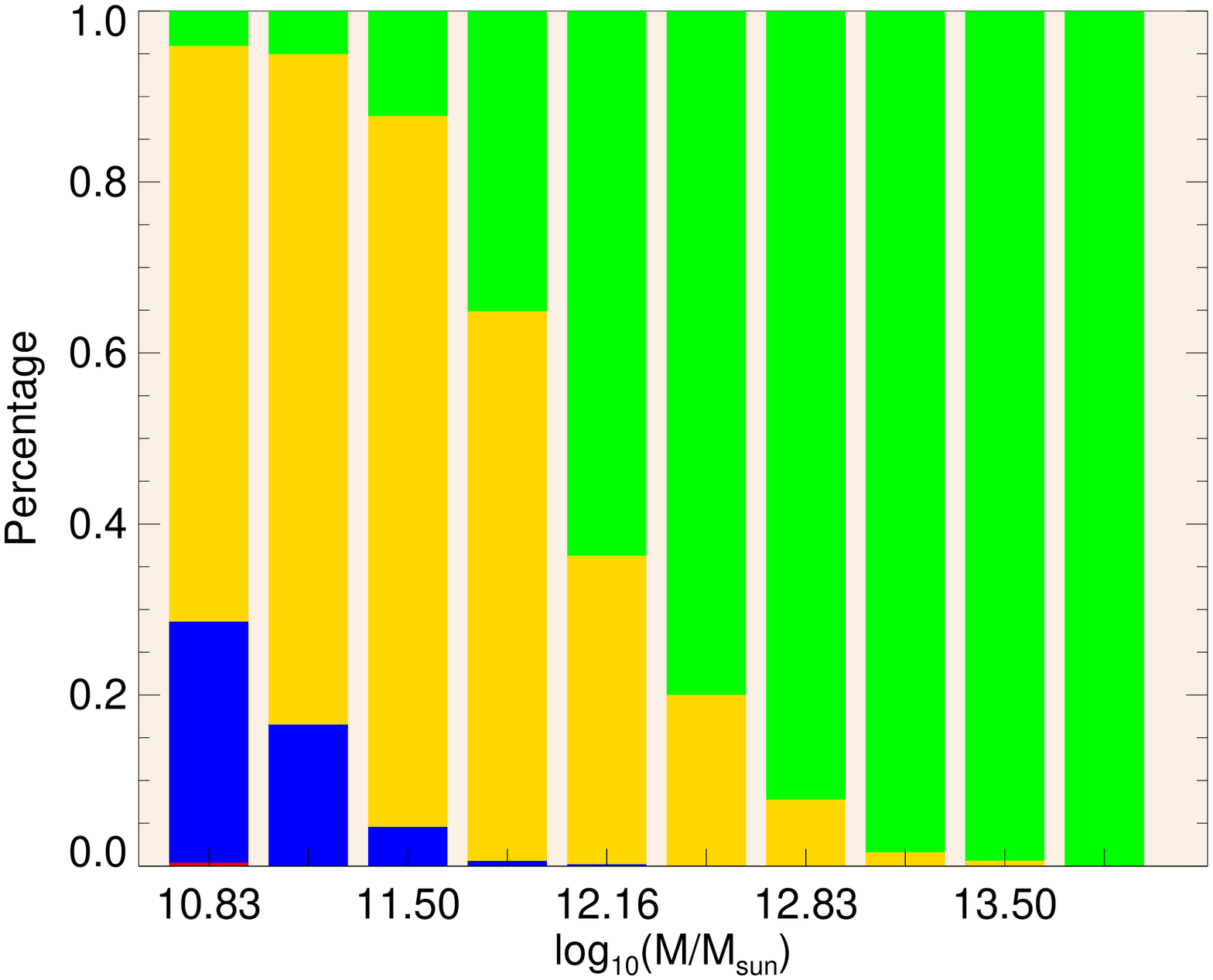}
\includegraphics[width=0.50\textwidth, trim=0 10 10 10, clip]{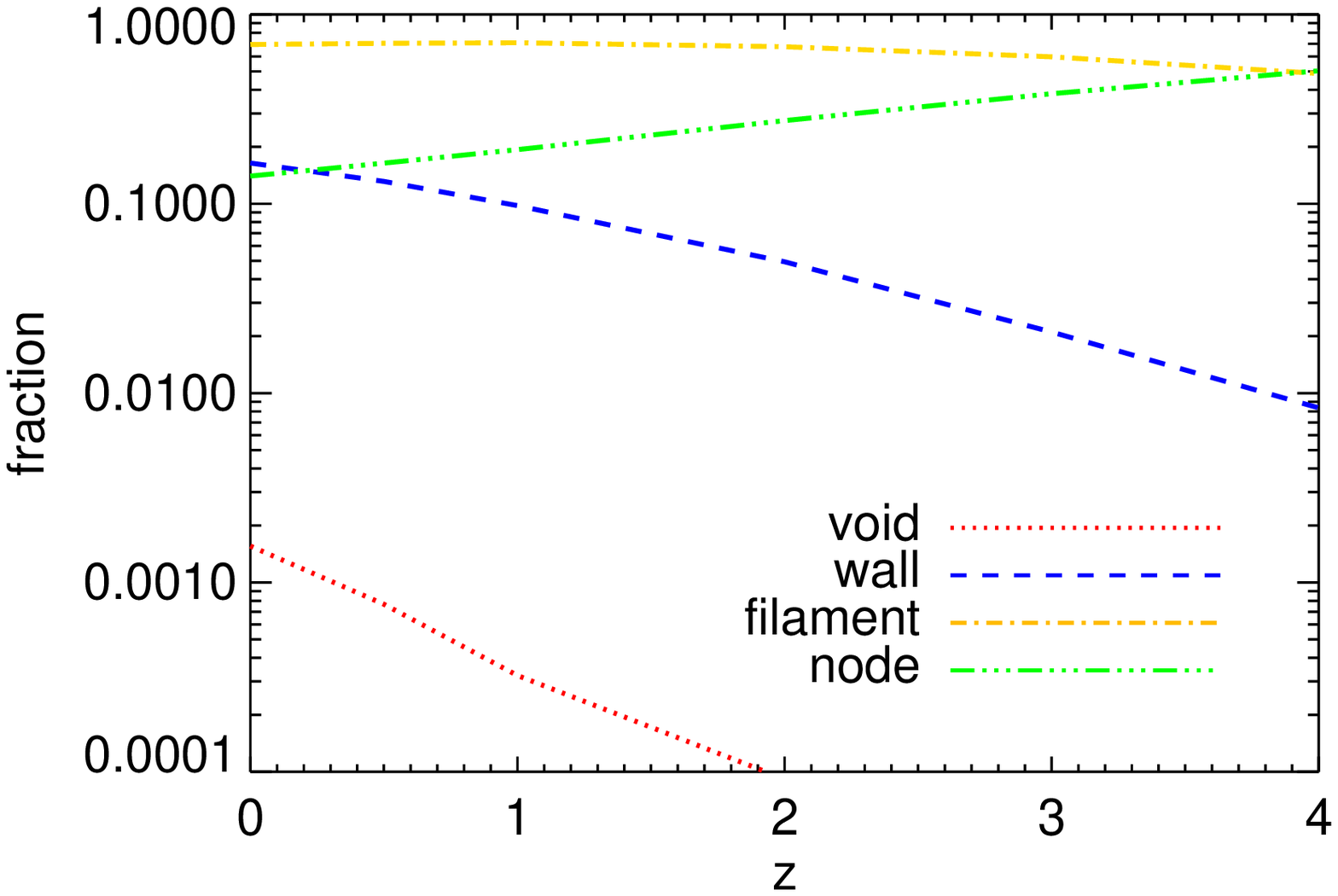}
\caption{Top: The cumulative halo mass function at $z=0.0$. Solid line indicates the result of all the halos, and dotted and dashed lines indicate mass function in different web environment. Middle: The fractions of halos in four types of environment in different halo mass bin, with bin size $10^{0.33}\  \rm{M_{\odot}}$, at $z=0.0$. Bottom: The fractions of halos in different cosmic web environment since $z=4.0$. }
\label{fig:halo_web}
\end{center}
\end{figure}

We use the friend-of-friend algorithm to search for dark matter halos in our simulation samples. The linking length parameter is 0.2. For the sake of reliability, only halos having more than 400 dark matter particles, corresponding to a mass of $4.1 \times 10^{10}\ \rm{M_{\odot}}$ are included in our analysis. The number of halos varies from 35000-65000 at $z<4$. The mass center and virial radius for each halo have been calculated. The virial radius is defined as the radius at which the volume-average density equals to 200 times of the critical density and is currently denoted as $R_{200}$, or $R_{vir}$. We also calculate the mass-averaged velocity within the virial radius, $\vec{v_c}$. 

The accumulative halo mass function at $z=0$ is presented as the solid black line in the top panel of Figure \ref{fig:halo_web}. For each halo, its environment is set to the environment type associated with its mass center. The distribution of halos in the cosmic web is also shown in Figure \ref{fig:halo_web}.  At $z=0$, about $70\%$ of the halos are residing in filaments. Meanwhile, a considerable fraction of halos less massive than $10^{11.0}\  \rm{M_{\odot}}$ live in walls and most of the halos massive than $10^{12.0}\ \rm{M_{\odot}}$ site in nodes. The bottom panel of Figure \ref{fig:halo_web} shows the fraction of halos hosted in different environment since $z=4.0$. It can be seen that most of halos reside in filaments since very early time. The fractions of halos living in nodes (walls ) decrease (increase) gradually after $z=4$. In comparison to \cite{2014MNRAS.441.2923C}, the fraction of halos residing in filaments is similar, but relatively more(less) halos are found in nodes (walls and voids) in our samples, which is likely due to the difference on web classification methods. As this work is only focusing on the halos in filaments, the selection of web classification would have a minor effect on the results. 

\subsection{gas accretion rate}

\begin{figure*}[htbp]
\begin{center}
%\epsscale{1.5}
\includegraphics[width=0.45\textwidth, trim=0 10 10 10, clip]{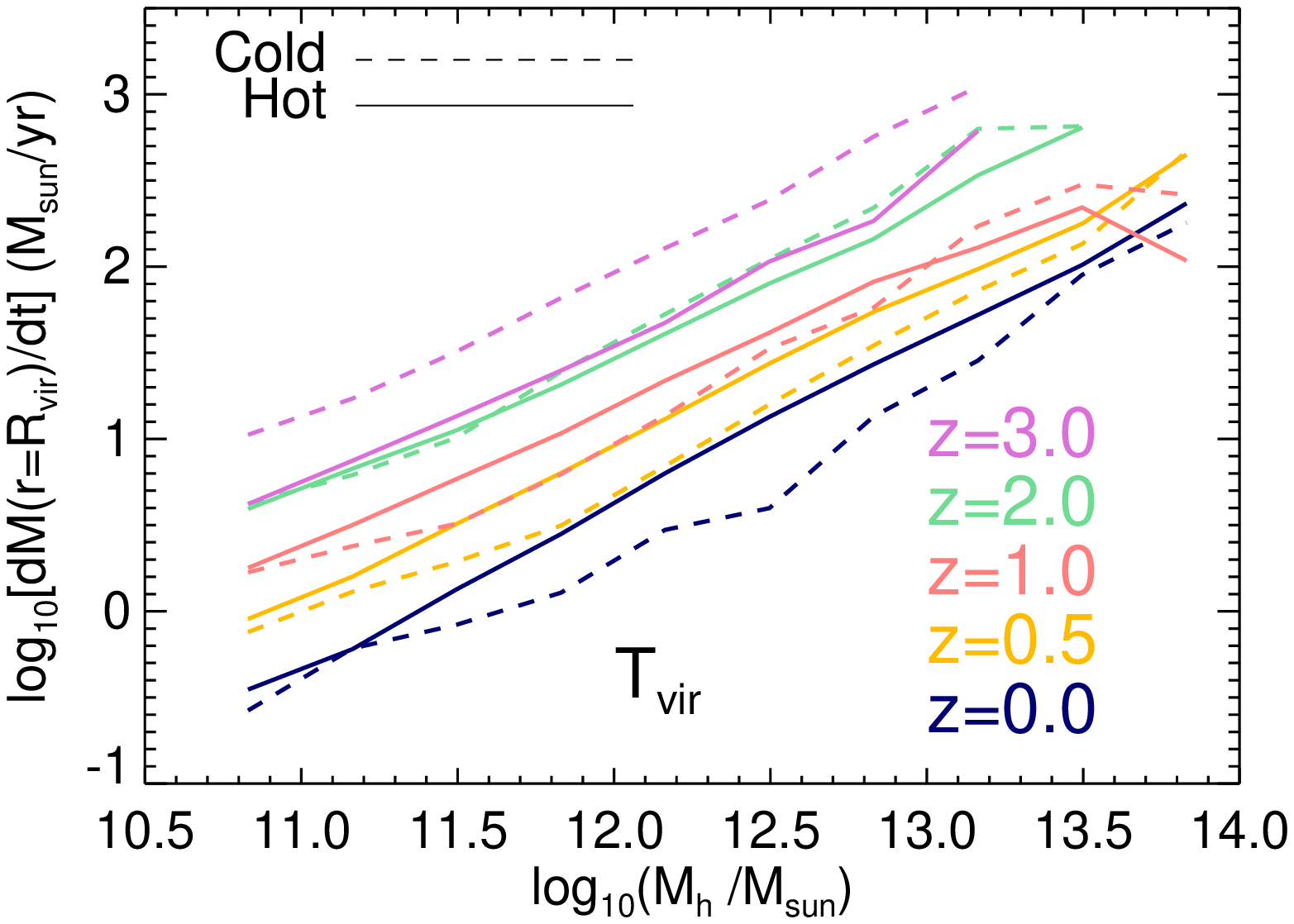}
\includegraphics[width=0.45\textwidth, trim=0 10 10 10, clip, ]{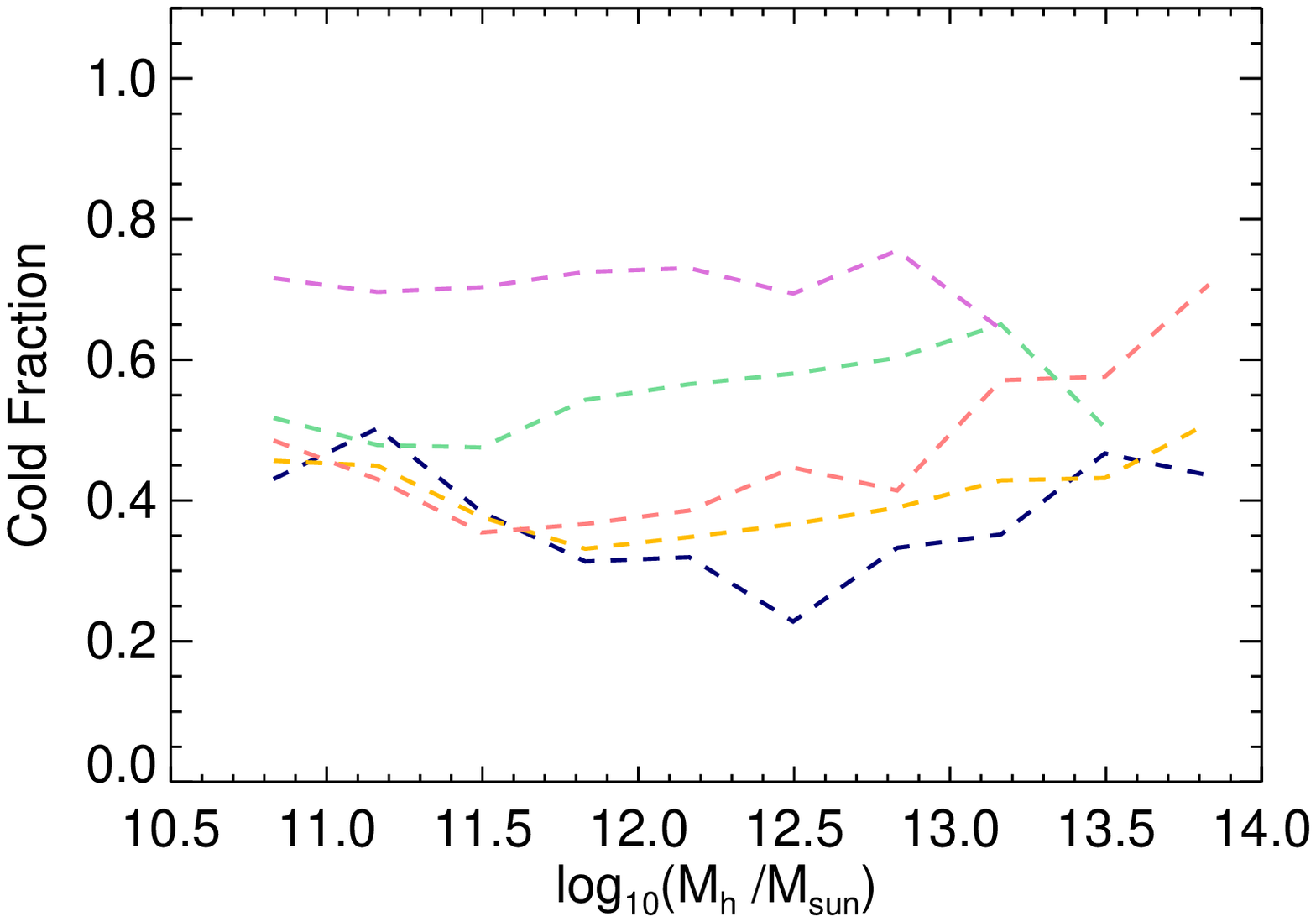}
\includegraphics[width=0.45\textwidth, trim=0 10 10 10, clip]{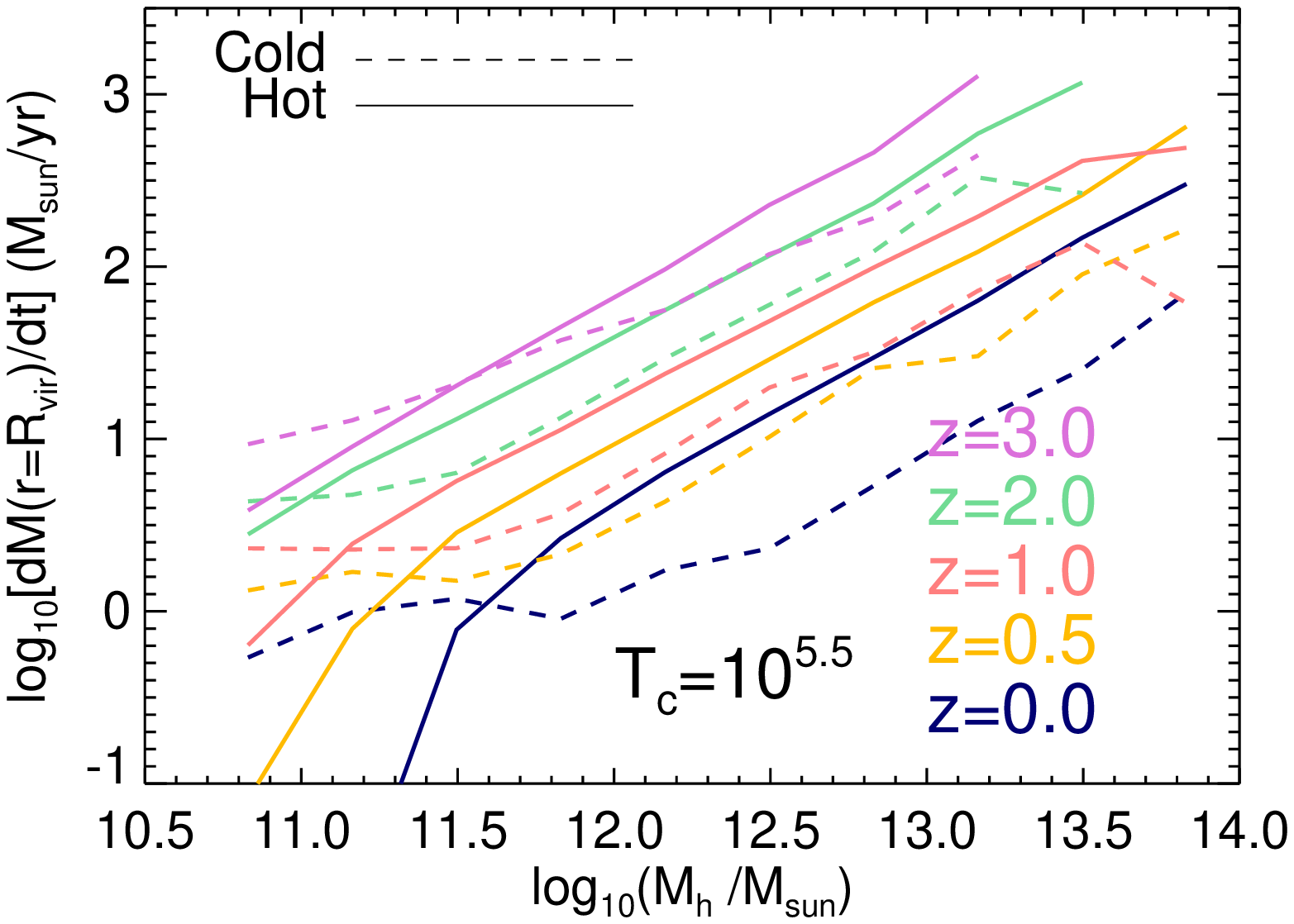}
\includegraphics[width=0.45\textwidth, trim=0 10 10 10, clip, ]{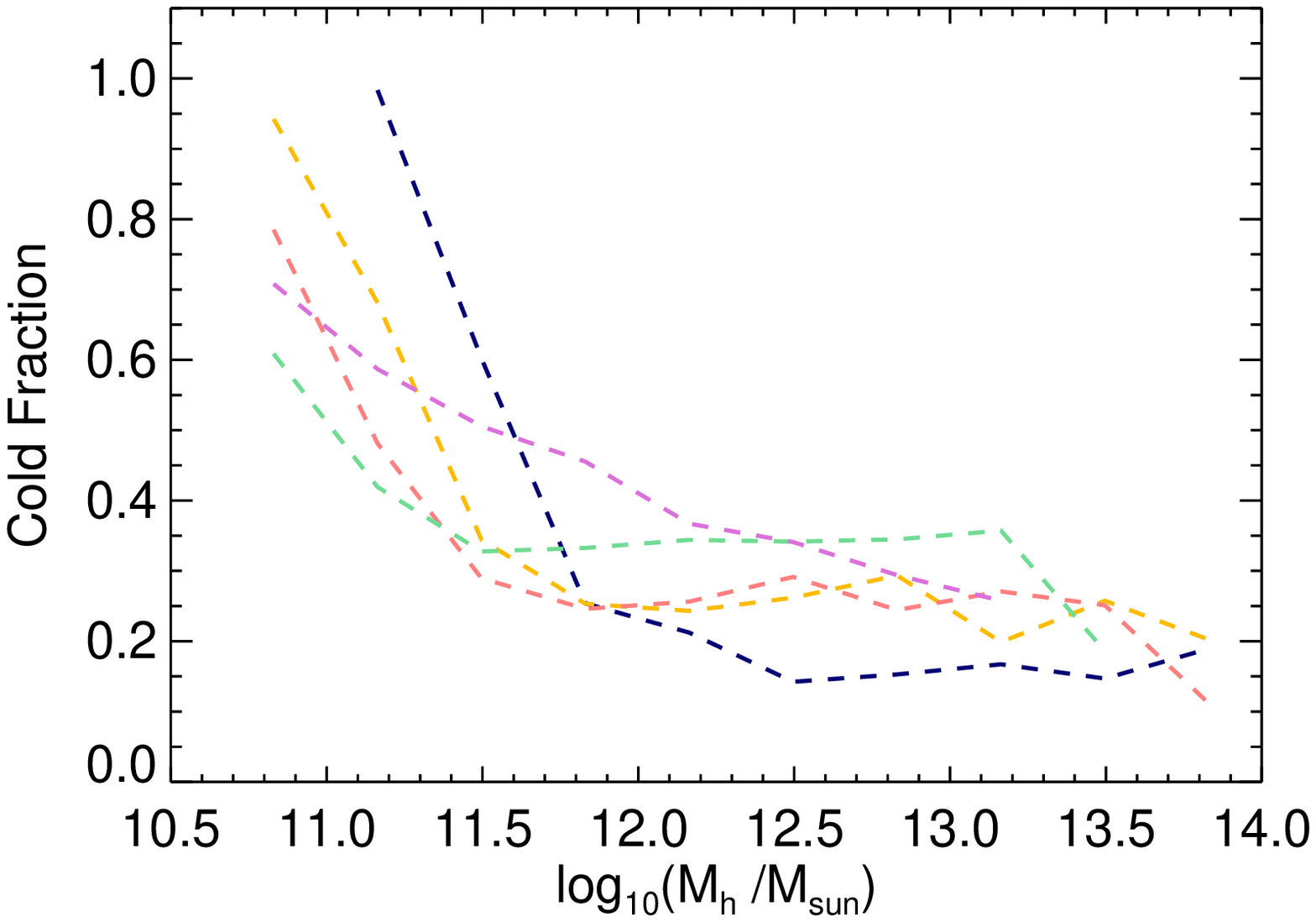}
\caption{Top Left: The mean gas accretion rate to halos at radius $r=R_{200}$ as a function of halo mass. Solid and dashed lines indicate the accreting rate of hot and cold modes respectively. The temperature threshold between cold and hot modes is set to the virial temperature of each halo.  Top Right: the corresponding cold fraction. Bottom row: same as the top row, but with a constant threshold temperature $10^{5.5}$ K. }
\label{fig:halo_acc}
\end{center}
\end{figure*} 

We use the similar method to \cite{2008MNRAS.390.1326O} to calculate the radial gas accretion rate onto dark matter halos, but with some minor variations. We use the halo's mass center as the reference point, instead of the baryonic density peak in the halo central region used by \cite{2008MNRAS.390.1326O}. For each halo, the gas properties are mapped from the AMR grid cells to concentric spherical shells centered at halo's mass center with various radius of $0.1R_{200}<r_s<1.0 R_{200}$. On the surface of each shell, we sample the density, $\rho$, temperature, $T$, and velocity $\vec{v}$ of gas at pixels with an angular resolution $3\ \rm{kpc}/r_s$, by smoothing the underlying fields in AMR grid cells. Then the total gas accretion rate at a shell surface at $r_s$ is defined as 

\begin{equation}
    \dot{M}(r_s)=\int{\rho (\vec{v}-\vec{v_c}) \cdot \vec{n} dS},
\label{eqn:acc}
\end{equation}
where $\vec{n}$ is the vector normal to the shell surface. 

In this work, we focus on the accretion rate at the virial radius, since a robust study on the gas accretion onto galaxies and its consequent effects on star formation will need simulations with more sophisticated sub-grid physics on star formation and feedback. Previous studies based on simulations have shown that gas accretion onto halos is bimodal, some gas being in a cold phase with temperature of below a few $10^5$ K and some being much hotter (\citealt{2005MNRAS.363....2K}; \citealt{2008MNRAS.390.1326O}; \citealt{2011MNRAS.414.2458V}; \citealt{2011MNRAS.417.2982F}). 
According to the gas temperature on shell surface, it is straightforward to measure the cold and hot gas accretion rate separately, simply by integrating over pixels with temperature above/below certain temperature threshold in Equation \ref{eqn:acc}. However, the cold accretion rate highly depends on the definition of ``cold'' and ``hot'', namely, the temperature threshold used to split two modes. Following previous studies (e.g. \citealt{2011MNRAS.417.2982F}; \citealt{2013MNRAS.429.3353N}), we use two definitions of temperature threshold between cold and hot modes, one is to place a constant threshold value, $10^{5.5}$ K, and the other is to adopt the virial temperature of each halo. The virial temperature is estimated as(e.g. \citealt{2001PhR...349..125B} ), 
\begin{equation}
    T_{\rm{vir}}=\frac{\mu m_p V_c^2}{2k_B} \simeq 3.5\times 10^5\ K\ (\frac{M_h}{10^{11}\rm{M_{\odot}}})^{2/3}(\frac{1+z}{3}),
\end{equation}
where $\mu$ is the mean molecular weight and equals approximately to $0.6$ for fully ionized primordial gas, and $V_c$ is the circular velocity at the virial radius.

\section{gas accretion rate onto halos} \label{sec:accretion}
We first investigate the gas accretion rate of all the halos in our sample, and compare the results to previous study. After that, we focus on the gas accretion rate tono halos in the environment of filaments to explore the impact of filaments. 

\subsection{overall results} 
The top left panel of Figure \ref{fig:halo_acc} shows the mean gas accretion rate onto halos at virial radius $r_s=R_{vir}$ as a function of halo mass between $z=3.0$ and $z=0.0$. The virial temperature of each halo is used as the threshold temperature between cold and hot accretion modes. The gas accretion rate increases with increasing halo mass and declines gradually with decreasing redshift. The variations of absolute gas accretion rate with the mass of the central halo and redshift are broadly consistent with many previous simulations (e.g. \citealt{2008MNRAS.390.1326O}; \citealt{2011MNRAS.417.2982F}; \citealt{2011MNRAS.414.2458V}). The top right panel presents the cold fraction, $f_c$, i.e., the ratio of accretion rate of gas with a temperature below virial temperature to the total gas accretion rate. Clearly, the cold fraction shows a weak dependence on halo mass, which agrees with results in \cite{2011MNRAS.417.2982F} and \cite{2013MNRAS.429.3353N}.  The cold fraction is around $70\%$ at $z=3$, and decrease to $50\%-60\%$ at $z=2$, and further to $\sim 40\%$ at $z \leq 0.5$. 

The bottom row in Figure \ref{fig:halo_acc} also presents the gas accretion rates onto halos in the cold and hot modes, as well as the corresponding cold fraction but with a constant temperature threshold $T=10^{5.5}$ K. We can see that the cold fraction decreases with increasing halo mass, which has been actually revealed by many previous studies (e.g. \citealt{2005MNRAS.363....2K}; \citealt{2008MNRAS.390.1326O}; \citealt{2011MNRAS.417.2982F};\citealt{2011MNRAS.414.2458V};\citealt{2013MNRAS.429.3353N}). The reason is that the temperature of gas around halos is on the order of the virial temperature, which depends on the halo mass (\citealt{2011MNRAS.414.2458V}; \citealt{2013MNRAS.429.3353N}). For halos with a virial temperature smaller than $10^{5.5}$ K, i.e., $ M_h \leq \sim 10^{11.5-12.0} \rm{M_{\odot}}$, most of the gas entering halos is cool than $10^{5.5}$ K, and hence are identified as cold accretion if the threshold temperature is set to $T=10^{5.5}$ K. 

\subsection{impact of filaments}

\begin{figure}[htbp]
\begin{center}
%\epsscale{1.5}
\includegraphics[width=0.48\textwidth, trim=0 10 10 10, clip]{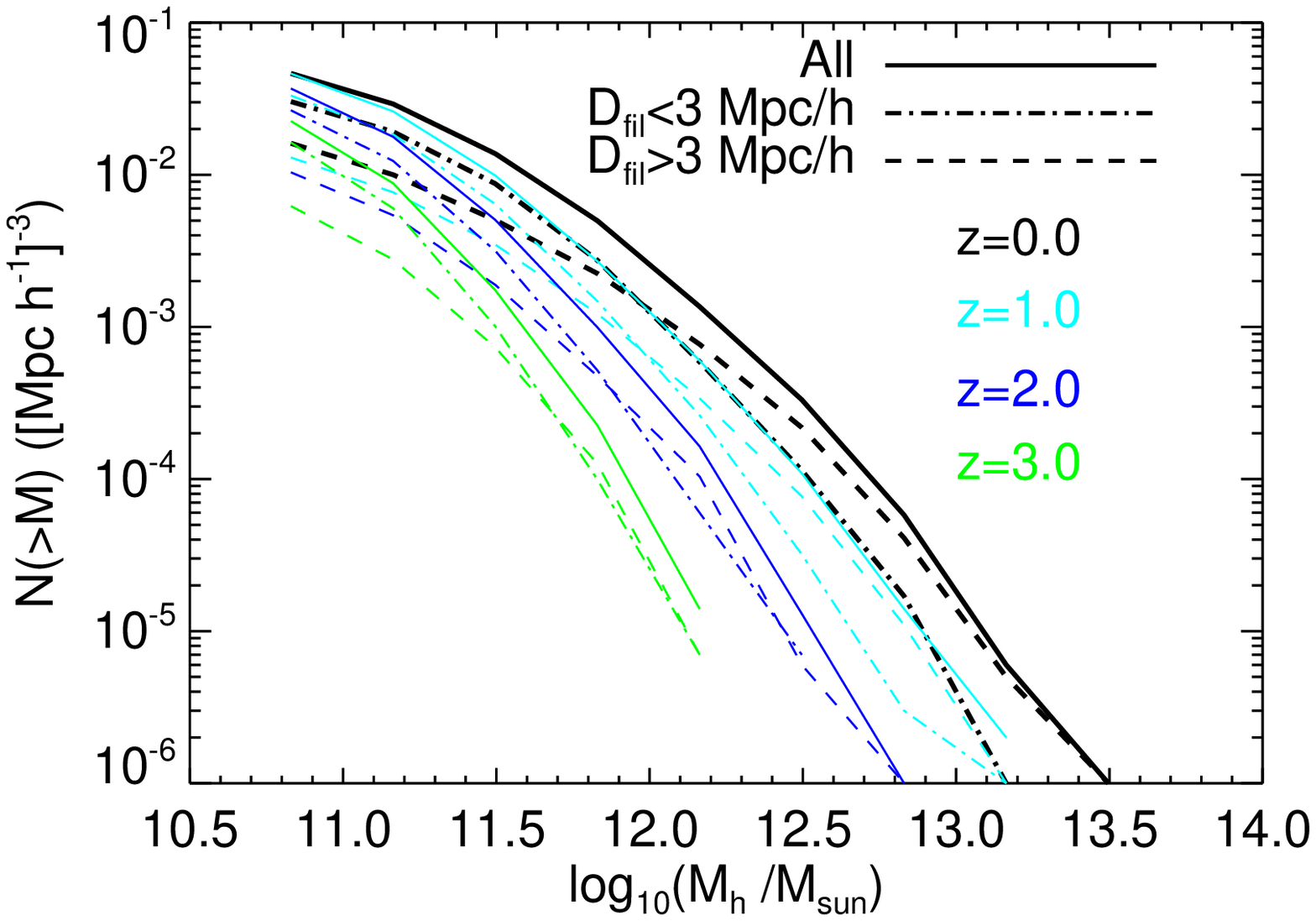}
\includegraphics[width=0.48\textwidth, trim=0 10 10 10, clip]{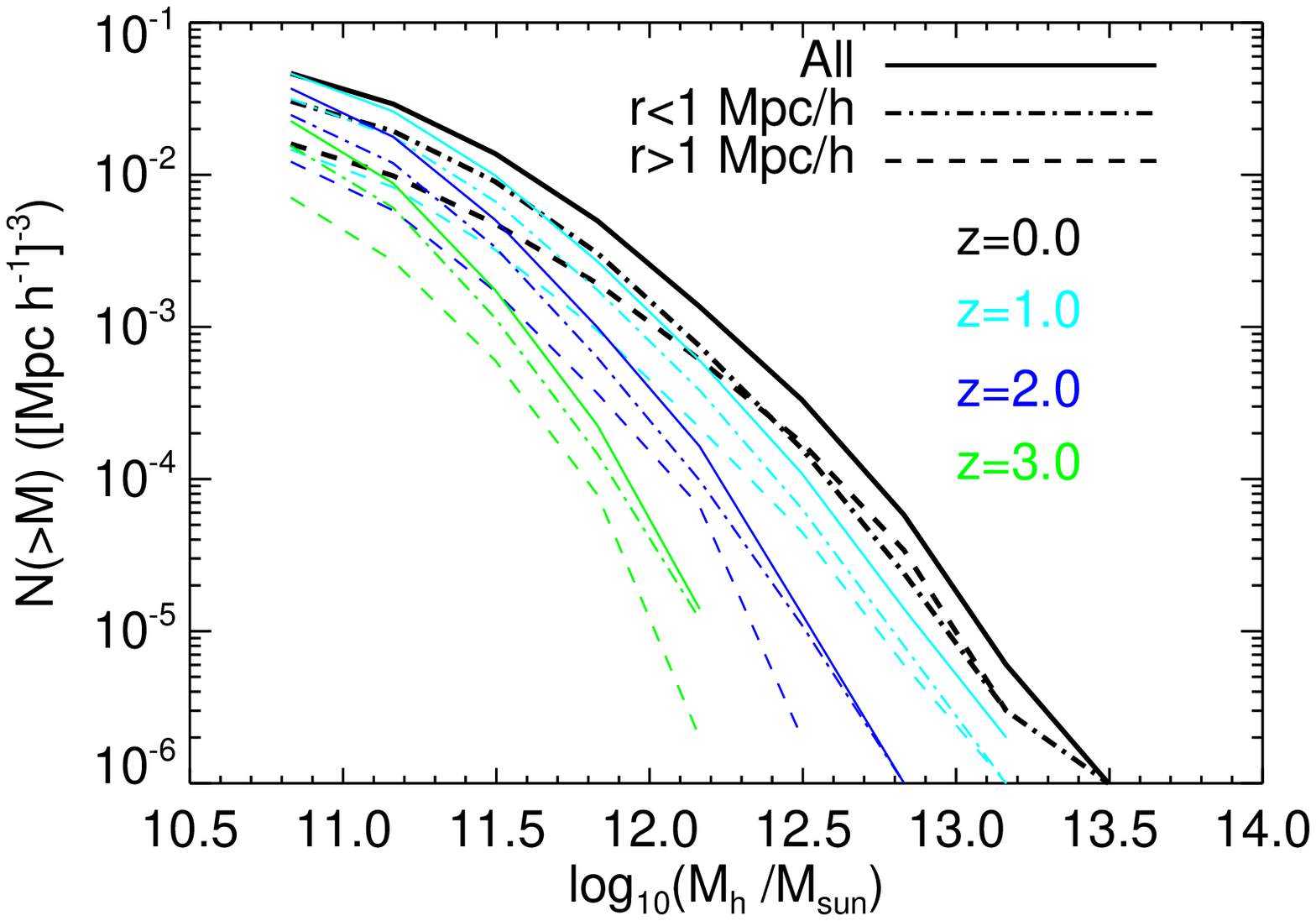}
\caption{The cumulative halo mass function in filaments as a function of halo mass. Top: Dotted and dashed lines indicate halos residing in relatively thin($D_{\rm{fil}}<3.0\ \rm{Mpc}/h$) and thick($D_{\rm{fil}} \geq 3.0\ \rm{Mpc}/h$) filaments respectively.  Bottom: Dotted and dashed lines indicate halos with a distance to the spine $r<1.0\ \rm{Mpc}/h$ and $r>1.0\ \rm{Mpc}/h$ respectively.} 
\label{fig:halo_fil}
\end{center}
\end{figure}

\begin{figure}[htbp]
\begin{center}
%\epsscale{1.5}
\includegraphics[width=0.38\textwidth, trim=10 10 10 10, clip, angle=90]{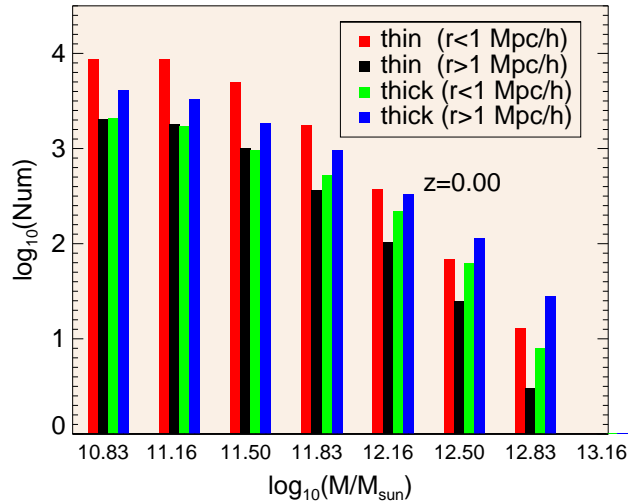}
\caption{The number of halos in different mass bins at $z=0$. The bin size is 0.33, the x axis shows the mid value in each bin. Halos residing in thin($D_{\rm{fil}}<3.0\ \rm{Mpc}/h$) filaments are divided into two subgroups with $r<1.0\ \rm{Mpc}/h$ (Red) and $r>1.0\ \rm{Mpc}/h$ (Black). Halos residing in thick($D_{\rm{fil}} \geq 3.0\ \rm{Mpc}/h$) filaments are divided into two subgroups with $r<1.0\ \rm{Mpc}/h$ (Green) and $r>1.0\ \rm{Mpc}/h$ (Blue).} 
\label{fig:halo_num_dept}
\end{center}
\end{figure}

To explore the environmental effect on the gas accretion rate to halos in filaments, we divide the halos residing in filaments into two categories according to two different methods respectively. The first method makes use of the local diameter of the filament segment where a halo is located. The threshold of local diameter is set to 3 $\rm{Mpc}/h$. As shown in \cite{2021ApJ...920....2Z}, for gas in filament segments with $\rm{D_{fil}} \geq 3\ \rm{Mpc}/h$, the typical temperature is higher than $10^{5.5}-10^{6.0}$ K. In contrast, most of the gas in filaments with $\rm{D_{fil}} \leq 3\ \rm{Mpc}/h$ would be cooler than $10^{5.5}-10^{6.0}$ K.  For the sake of briefness, filaments with $\rm{D_{fil}} \geq 3\ \rm{Mpc}/h$ will be referred to as `thick/prominent'', and filaments with $\rm{D_{fil}} \leq 3\ \rm{Mpc}/h$ as ``thin/tenuous" in the following paragraphs. The top panel in Figure \ref{fig:halo_fil} shows the cumulative halo mass function in  thick and thin filaments, as well as in all of the filaments  at different redshifts since $z=3.0$. About $ 65\%$ of halos in filaments are found in tenuous filaments with $\rm{D_{fil}} \leq 3\ \rm{Mpc}/h$ at $z=0$. Nevertheless, halos massive than $10^{12} \ \rm{M_{\odot}}$ are more likely to be hosted by thick filaments. The fractions of halos in thick filaments decline as redshift \textbf{increases}. This is mainly because the number frequency of thick filaments decreases with increasing redshift , which has been demonstrated by visual expression in previous study and by quantitative analysis in \cite{2021ApJ...920....2Z}. 

The second splitting method is to characterize the filament effect by using the perpendicular distance to the filament spine from the halo center. We take a typical value of distance of 1 $\rm{Mpc}/h$ as the dividing line between two categories of halos. The bottom panel of Figure \ref{fig:halo_fil} shows the cumulative mass function of halos in each of two categories. We find that about two thirds of the halos in filaments have a distance smaller than 1 $\rm{Mpc}/h$ from the spine at $z \leq 3.0$. For a more straightforward view, Figure \ref{fig:halo_num_dept} shows the numbers of halos in different mass bins according to two different splitting methods at $z=0$. Thick filaments host more halos with $r \geq \rm{Mpc}/h$ than thin filaments. Only around a dozen of halos are massive than $10^{13.0}\ \rm{M_{\odot}}$ in our simulation, and they not shown in Figure \ref{fig:halo_num_dept}.

\begin{figure}[htbp]
\begin{center}
%\epsscale{1.5}
\hspace{-0.0cm}
\includegraphics[width=0.48 \textwidth, trim=0 10 10 10, clip]{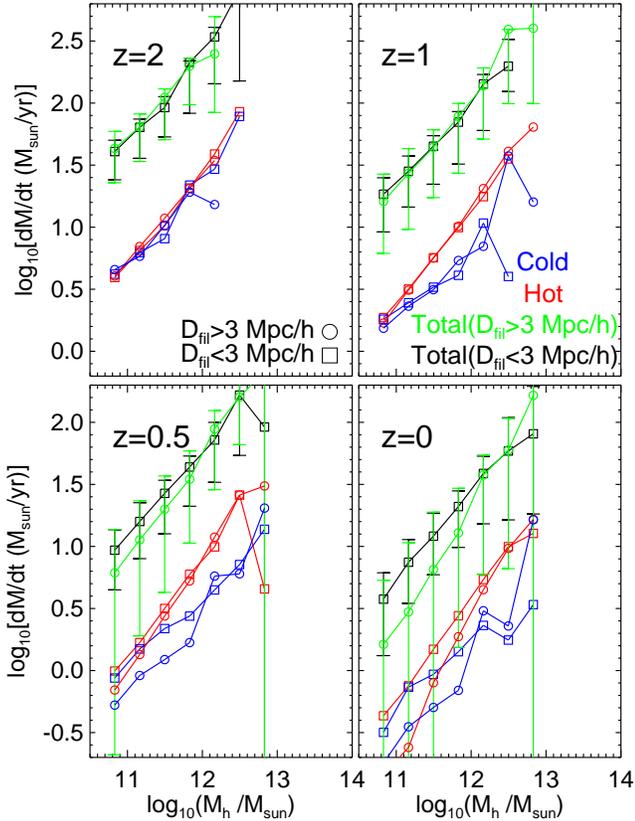}
\caption{The mean gas accretion rate to halos in filaments as a function of halo mass between $z=2$ and $z=0$. Halos are divided into two groups according to the local diameter $\rm{D_{fil}}$. Square and circle indicate halos hosted by `thick' filaments with $\rm{D_{fil}} \geq 3\ \rm{Mpc}/h$ and `thin' filaments with $\rm{D_{fil}} \geq 3\ \rm{Mpc}/h$ respectively. Black, and green lines indicate the mean value of the total accretion rate of halos in thin and thick filaments respectively. The upper and lower bars show the 75th and 25th percentiles
in each bins.  The total accretion rate have been shifted upward by 0.7 dex, for the sake of clarity. Red and blue lines indicate the mean value of  hot and cold accretion rate.  The virial temperature of each halo is adopted as the threshold gas temperature between hot and cold accretion modes.}
\label{fig:fil_acc_dia}
\end{center}
\end{figure}

\begin{figure}[htbp]
\begin{center}
%\epsscale{1.5}
\hspace{-0.0cm}
\includegraphics[width=0.48\textwidth, trim=0 10 10 10, clip]{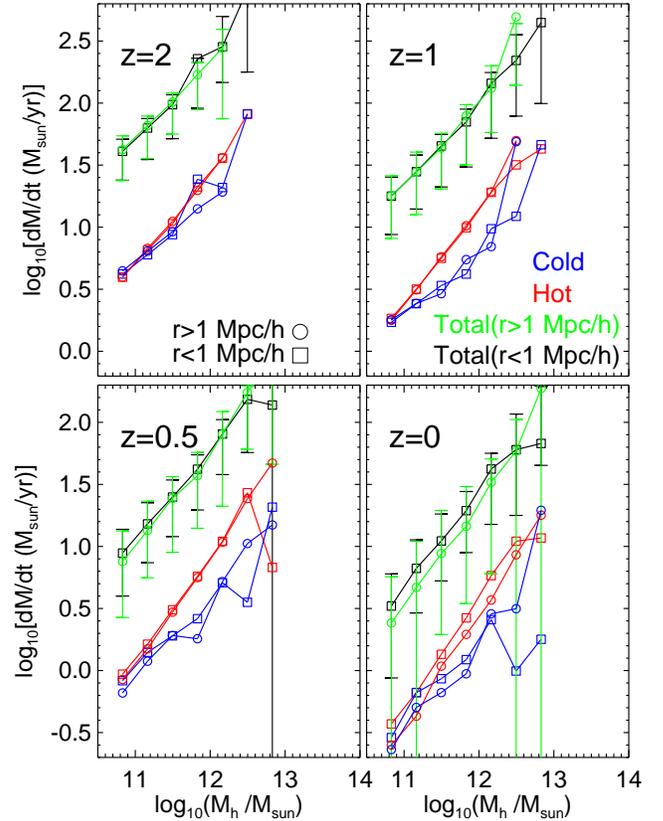}
\caption{Same as Figure \ref{fig:fil_acc_dia}, but halos in filaments are divided into two groups according to the distance to the spine.}
\label{fig:fil_acc_dep}
\end{center}
\end{figure}

Figure \ref{fig:fil_acc_dia} compares the gas accretion rate of two categories of halos residing in filaments with width $\rm{D_{fil}}$ smaller and larger than $3.0\  \rm{Mpc}/h$ respectively since $z=2$. At redshifts $z \geq 1.0$, the difference between two groups \textbf{are} also negligible, while down to redshift $z=0.5$, the gas accretion rate onto halos less massive than $M_h=10^{12.0}\ \rm{M_{\odot}}$ exhibits moderate difference between them. The discrepancy become more evident for less massive halos. The difference of accretion rate between the two groups in cold mode is stronger than in hot mode, where the virial temperature is used as the division between two modes. Halos residing in filaments segments with widths smaller than $3.0\ \rm{Mpc}/h$ have higher gas accretion rates than those in filaments with width larger than $3.0\ \rm{Mpc}/h$ by about $20\%$ at $\rm{M_h \sim 10^{11}\ M_{\odot}}$. The differences of mean gas accretion rate between two groups of halos can be as large as a factor of 2-3 at $z=0.0$ for halos with $\rm{M_h}<10^{12.0}\ \rm{M_{\odot}}$. The upper and lower bars in Figure \ref{fig:fil_acc_dia} show the 75th and 25th percentiles, of total accretion rate in each bins. The distributions indicate that a considerable fraction of less massive halos in thick filaments are acquiring gas at a rate much lower than their counterparts in thin filaments at $z=0$. 

At the present time, the differences between two groups of halos in cold and hot modes are comparable to each other for halos less massive than $10^{12.0}\ \rm{M_{\odot}}$. For halos more massive than $10^{12.3}\ \rm{M_{\odot}}$, there are notable fluctuations on the accretion rate in both thick and thin filaments, and the fluctuations are wilder in more massive bins. These fluctuations at the high mass end should mainly be due to a very limited number of halos . Accordingly, the results on halos more massive than $10^{12.3}\ \rm{M_{\odot}}$ will be not included in the following discussion.

Figure \ref{fig:fil_acc_dep} shows the gas accretion rate onto two groups of halos in filaments as a function of halo mass, in which the second group splitting method has been applied. There is barely any difference between two groups of halos that have different distances to the spine, denoted as r, at $z \geq 0.5$. At $z=0$, the gas accretion rate onto halos far from the spine with $r>1\ \rm{Mpc}/h$ is lower than halos closer to the spine by $\sim20\%$ in the halo mass range $10^{10.8-12.0}\ \rm{M_{\odot}}$. The decline is similar for both hot and cold accretion. At the first glance, these features at $z=0$ are somewhat inconsistent with many previous studies which have shown that the star formation activity of galaxies tends to decrease, and hence indicate suppressed gas supply, as the distance to filaments spine decreases (e.g. \citealt{2016MNRAS.457.2287A}; \citealt{2017MNRAS.466.1880C}; \citealt{2017A&A...600L...6K}; \citealt{2018MNRAS.474..547K}; \citealt{2020A&A...638A..75B}; \citealt{2021arXiv210513368W}). 

The probable explanation is as follows. In our simulation,  the number of halos with a distance to the spine $r \geq 1\  \rm{Mpc}/h$ in thick filaments are larger than that in thin filaments. Therefore, given that halos in thick filaments have lower gas accretion rate than halos in thin filaments at redshift below $1.0$, halos with a distance to the spine of $r>1\ \rm{Mpc}/h$ would tend to have a lower gas accretion rate than those halos with $r<1\ \rm{Mpc}/h$. 

\begin{figure}[htbp]
\begin{center}
%\epsscale{1.5}
\includegraphics[width=0.38\textwidth, trim=10 10 10 10, clip, angle=90]{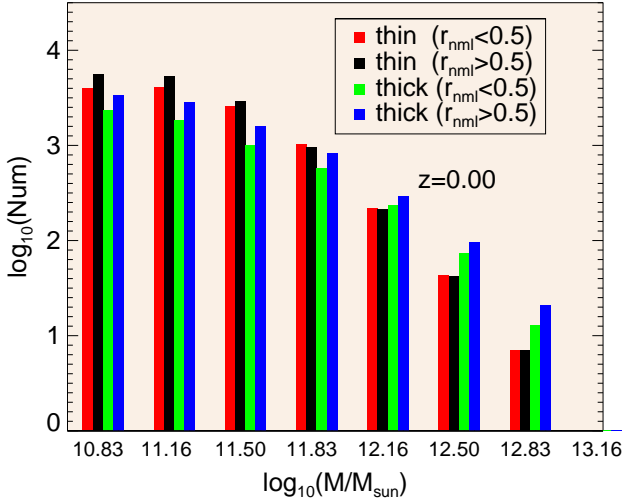}
\caption{The number of halos in different mass bins at $z=0$. Halos residing in thin filaments are divided into two subgroups with the res-caled distance to filament spine $r_{nml}<0.5$ (Red) and $r_{nml}>0.5$ (Black). Halos residing in thick filaments are also divided into two subgroups with $r_{nml}<0.5$ (Green) and $r_{nml}>0.5$ (Blue).} 
\label{fig:halo_num_dnml}
\end{center}
\end{figure}

\begin{figure}[htbp]
\begin{center}
%\epsscale{1.5}
\hspace{-0.0cm}
\includegraphics[width=0.48 \textwidth, trim=0 10 10 10, clip]{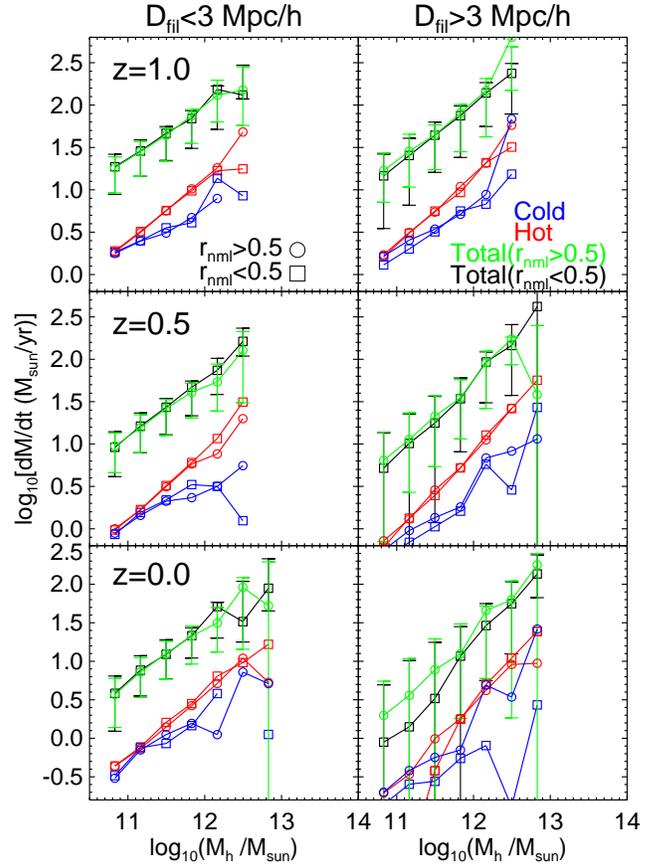}
\caption{The mean gas accretion rate to halos in filaments as a function of halo mass between $z=1$ and $z=0$. Halos in filaments are divided into two groups with the local diameter of filaments $\rm{D_{fil}}<3\ \rm{Mpc}/h$ (Left) and $\rm{D_{fil}}>3\ \rm{Mpc}/h$ (Right) respectively. Each group is separated to two subgroup, shown with open squares and circles, according to the halo's rescaled distance to the filament spine $\rm{r_{nml}}$. Meaning of different colors are similar to Figure \ref{fig:fil_acc_dia}. The upper and lower bars show the 75th and 25th percentiles of total accretion rate in each bins. The total accretion rate have been shifted upward by 0.7 dex, for the sake of clarity.}
\label{fig:fil_acc_dnm}
\end{center}
\end{figure}

We further separate each of the two groups of halos residing in thick and thin filaments into two subgroups according to whether a halo's re-scaled distance to the spine of filaments, $\rm{r_{nml}=r/R_{fil}}$, is smaller than 0.5 or not. Here, $\rm{R_{fil}=D_{fil}/2}$ is the local radius of filaments. By this splitting measure, we aim to exclude the effect of filament width and probe whether there is any difference between halos in the inner and outer region of filaments. Figure \ref{fig:halo_num_dnml} presents the number of halos in four subgroups at $z=0$. Generally, the number of halos in outer region is larger than that in the inner region. 

Figure \ref{fig:fil_acc_dnm} shows the gas accretion rate onto halos in four subgroups since $z=1.0$. We can see that halos in the inner and outer regions of filaments are assembling gas in the same rate till $z=0.0$ in thin filaments, and till $z=0.5$ in thick filaments. At $z=0$, for halos in thick filaments and with mass lower than $10^{11.6}\ \rm{M_{\odot}}$, the mean gas accretion rate of halos in the inner region are lower than their counterparts in the outer region by $\sim 50\%$. We will discuss the probable reason in the next subsection. 

Our results could provide an explanation to the recent observational work by \cite{2021ApJ...906...68L}, in which they did not find a clear gradient of HI fraction for galaxies in filaments around the Virgo cluster along the direction perpendicular to the filament spine. Galaxies samples with a distance to the filament spine less than 3.5 times of the scale length of filaments are used in their work. Note that most of the galaxies in \cite{2021ApJ...906...68L} are hosted by relatively thin filaments, with scale length $R_s$ smaller than 0.5 Mpc/h. These filaments would have minor effects on the properties of halos and galaxies, given our findings presented in this section. 

\section{gas distribution in filaments}
The suppressed gas accretion rate found for halos residing in thick filaments at $z \leq 0.5$ may be related to the properties of the intergalactic gas locating in thick filaments. In this section, we first explore the spatial distribution of thick and thin filaments and then analyze the density and thermal properties of gas in the cosmic web, especially in two groups of filaments with different widths. 

\begin{figure}[htbp]
\begin{center}
%\epsscale{1.5}
\hspace{-0.0cm}
\includegraphics[width=0.5 \textwidth, trim=60 40 40 30, clip]{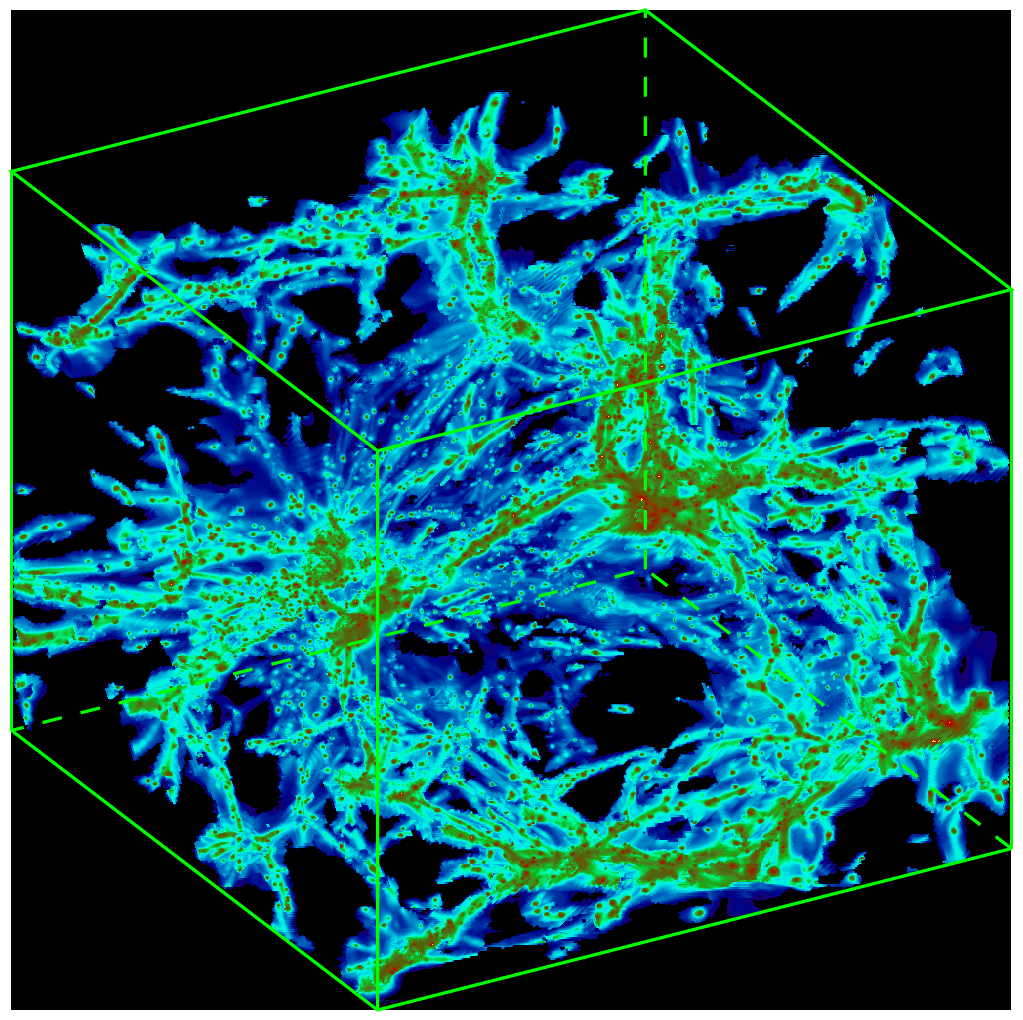}
\includegraphics[width=0.5 \textwidth, trim=60 40 40 30, clip]{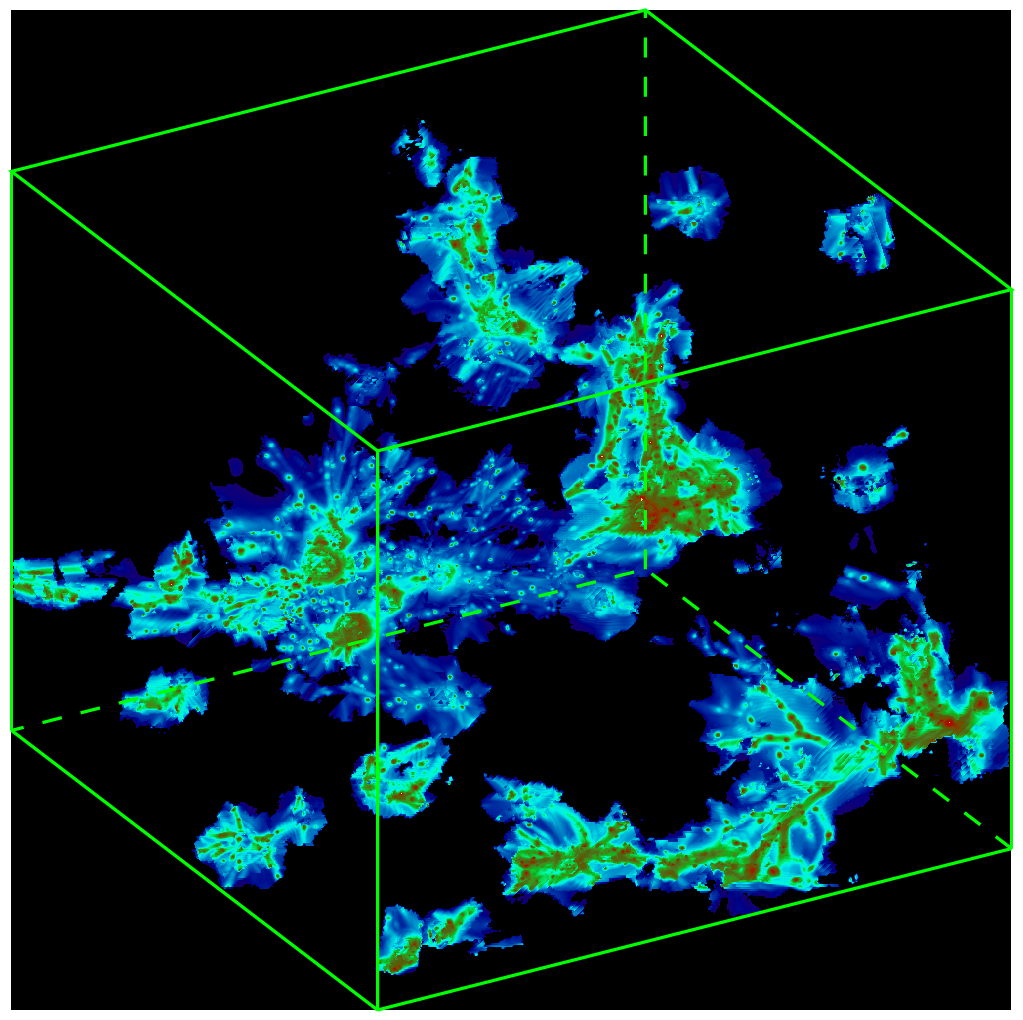}
\caption{Top: The gas distribution in filaments in a cubic box of volume $(33.3\ \rm{Mpc}/h)^3$ at $z=0.0$. Bottom: The gas distribution in filaments segments with local diameter $\rm{D_{fil}>3.0\ Mpc}/h$.}
\label{fig:den_fila}
\end{center}
\end{figure}

In the top panel of Figure \ref{fig:den_fila}, we can see the gas distribution in filaments in a sub volume of the simulation box at $z=0$. The bottom panel of Figure \ref{fig:den_fila} shows the gas distribution in thick filaments segments with local diameter $\rm{D_{fil}>3.0\ Mpc}/h$ in the same volume. Thick filaments are usually found in high-overdensity regions, consistent with the expectation and previous works (e.g. \citealt{2014MNRAS.441.2923C}). Meanwhile, thin filaments with $\rm{D_{fil}<3.0\ Mpc}/h$ usually appear in the middle ground between underdense voids and high-overdensity regions. 

\begin{figure}[htbp]
\begin{center}
%\epsscale{1.5}
\hspace{-0.0cm}
\includegraphics[width=0.48 \textwidth, trim=0 10 10 10, clip]{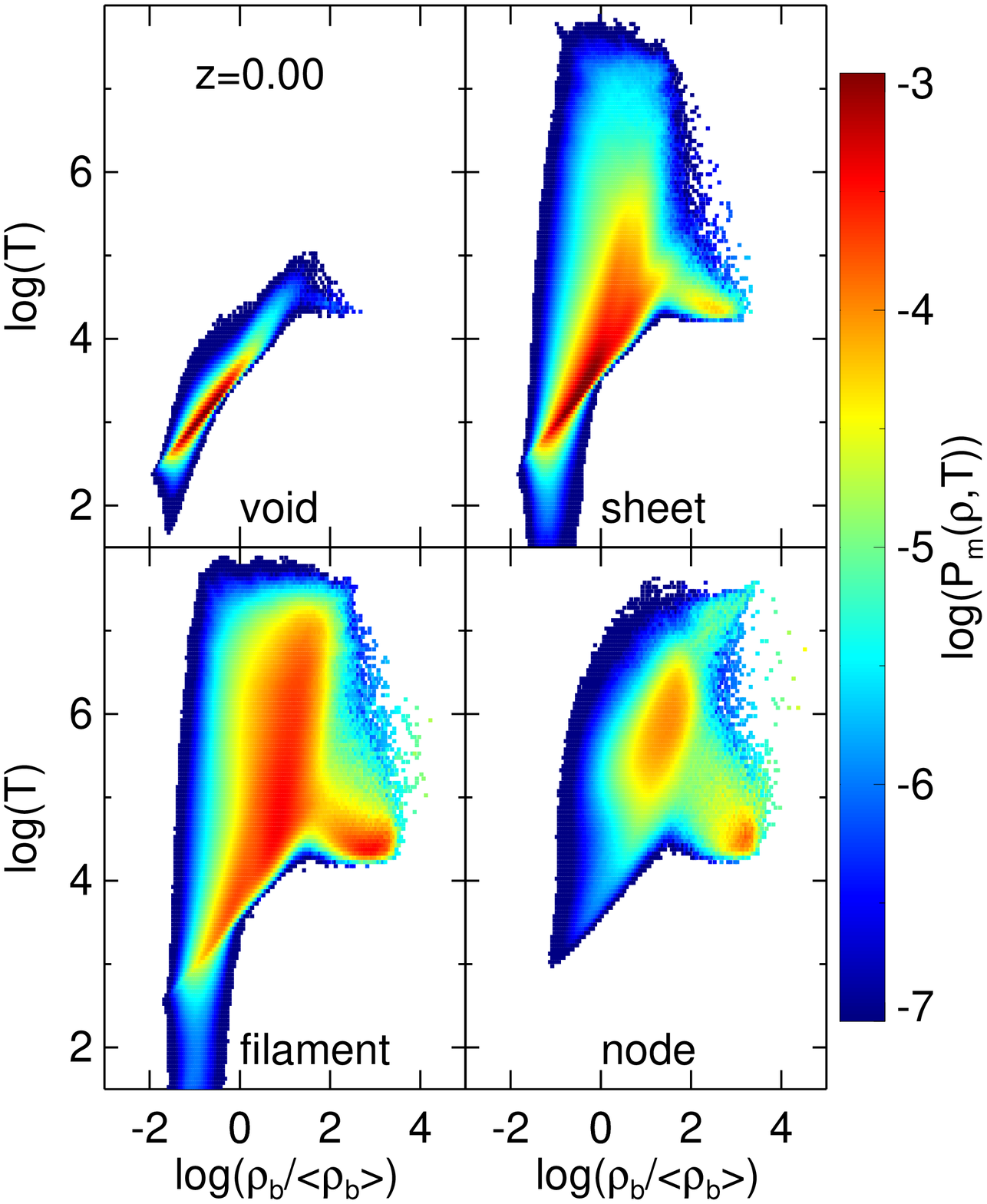}
\caption{The mass weighted distribution of intergalactic medium in density-temperature phase space at redshift 0. Top-left, top-right, bottom-left and bottom-right indicates the IGM in the environment of voids, walls, filaments and nodes respectively. Gas within the virial radius of dark mass halos is excluded in this plot.}
\label{fig:pdftem_web}
\end{center}
\end{figure}

\begin{figure}[htbp]
\begin{center}
%\epsscale{1.5}
\hspace{-0.0cm}
\includegraphics[width=0.48 \textwidth, trim=0 0 10 10, clip]{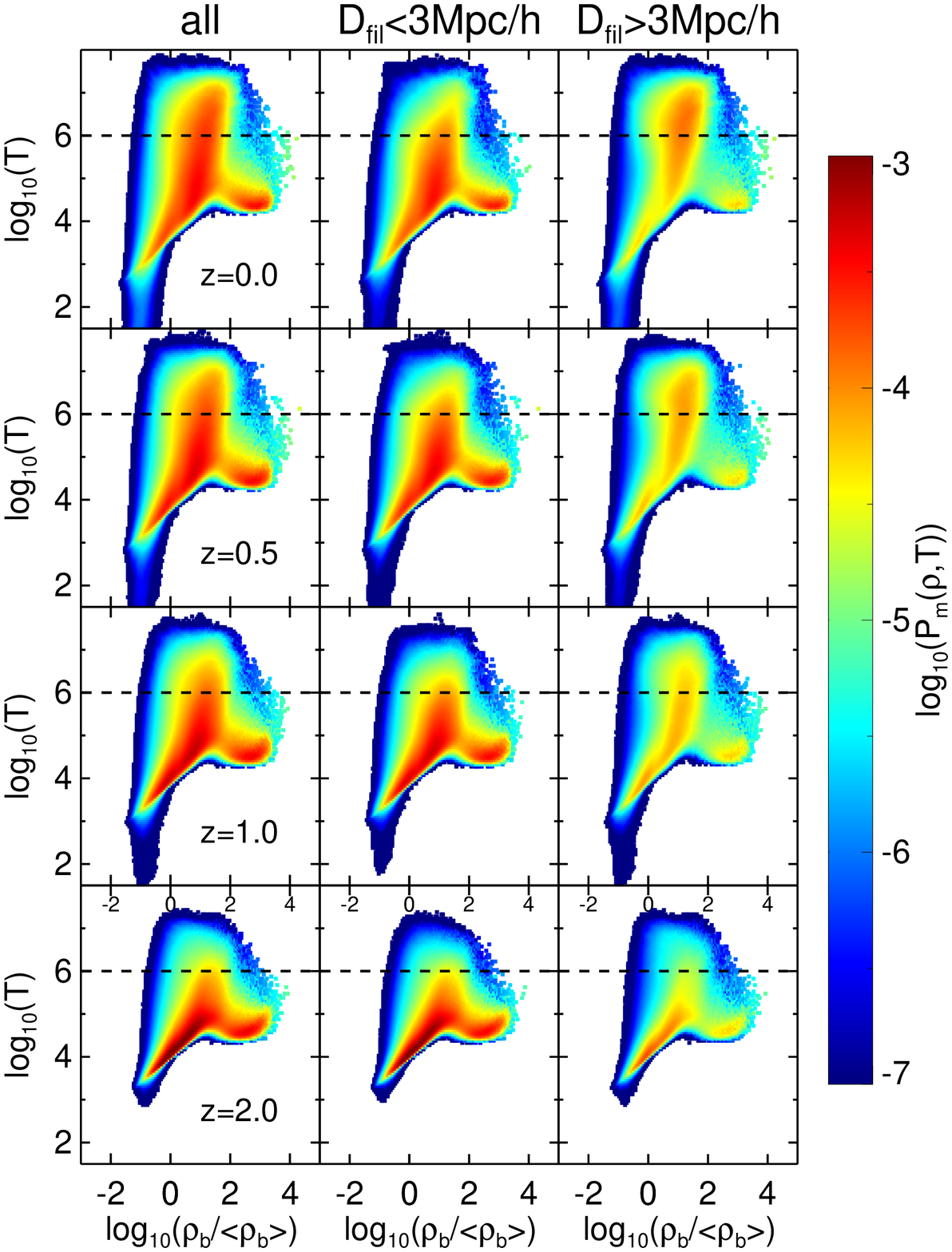}
\caption{Left: The mass weighted distribution of intergalactic medium residing in filaments in density-temperature phase space at redshift 0 (top), 0.5 (second row), 1.0 (third row) and 2.0 (bottom). The middle and right column show the IGM in filaments with $\rm{D_{fil}<3\ Mpc}/h$ and $\rm{D_{fil}>3\ Mpc}/h$ respectively. Gas within the virial radius of dark mass halos is excluded in this plot }
\label{fig:pdftem_zevo}
\end{center}
\end{figure}

Figure \ref{fig:pdftem_web} illustrates the distribution of intergalactic gas in the density-temperature space at $z=0$. Note that, gas within the virial radius of dark matter halos are excluded in the results shown in Figure \ref{fig:pdftem_web}. The distributions of gas located in four types of web environment are indicated in four panels respectively. The gas in both thin and thick filaments have been accounted for. Our results are generally consistent with previous studies based on various simulations such as Illustris and IllustrisTNG (e.g. \citealt{2016MNRAS.457.3024H}; \citealt{2019MNRAS.486.3766M}). A considerable fraction of the IGM is residing in filaments, typically with overdensity $\sim 1-100$ and temperature $10^{4}-10^{8}$ K. Moreover, most of the Warm-Hot intergalactic medium (with temperature $10^{5.5}-10^{7}$ K) are contained in filaments. The typical gas temperature in voids and walls is below $10^{5}$ K, which is cooler than gas in filaments.

Figure \ref{fig:pdftem_zevo} shows the evolution of gas phase in filaments between $z=2.0$ and $z=0.0$. The left column presents the evolution of all the intergalactic gas in filaments, while the middle and right columns present the evolution of gas in thin filaments with $\rm{D_{fil}<3\ Mpc}/h$ and thick filaments with $\rm{D_{fil}>3\ Mpc}/h$ respectively. We can see that most of the gases in filaments are cooler than $10^{6}$ K at $z=2.0$. Thereafter, the mass fractions of gases hotter than $10^{6}$ K grow gradually with the cosmic time, which should be mainly caused by gravitational collapse heating. The increase of gas fraction with $T>10^6$ K is dominated by thick filaments with $\rm{D_{fil}>3\ Mpc}/h$. A considerable fractions of gas residing in the filaments with $\rm{D_{fil}>3\ Mpc}/h$ have become hotter than $T>10^6$ K since $z=1.0$. While for filaments with $\rm{D_{fil}<3\ Mpc}/h$, the fraction of gas hotter than $10^6$ K remains small even at $z=0.0$.

\begin{figure*}[htbp]
\begin{center}
%\epsscale{1.5}
\hspace{-0.2cm}
\includegraphics[width=0.48 \textwidth, trim=0 10 10 10, clip]{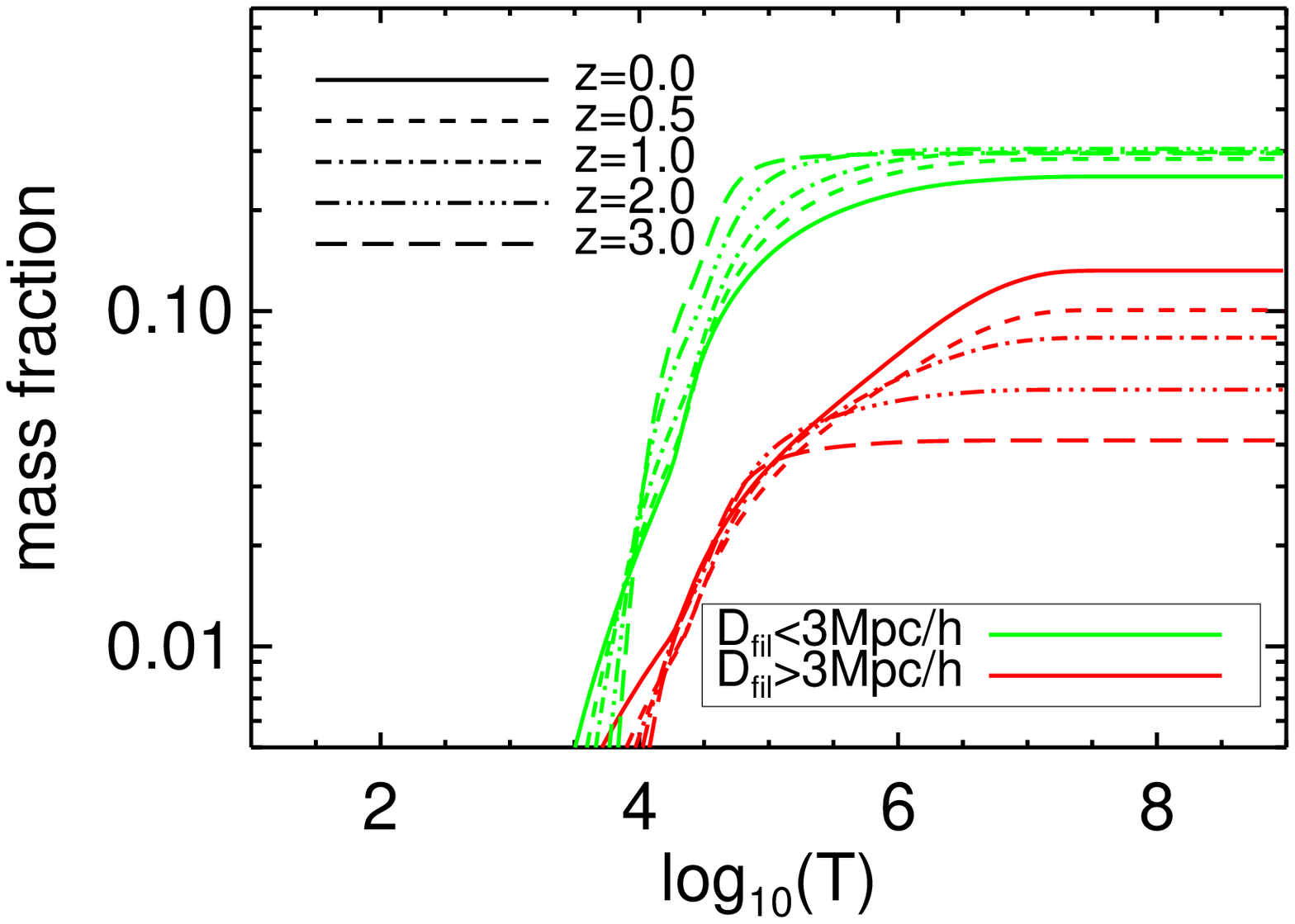}
\hspace{-0.0cm}
\includegraphics[width=0.48  \textwidth, trim=0 10 10 10, clip]{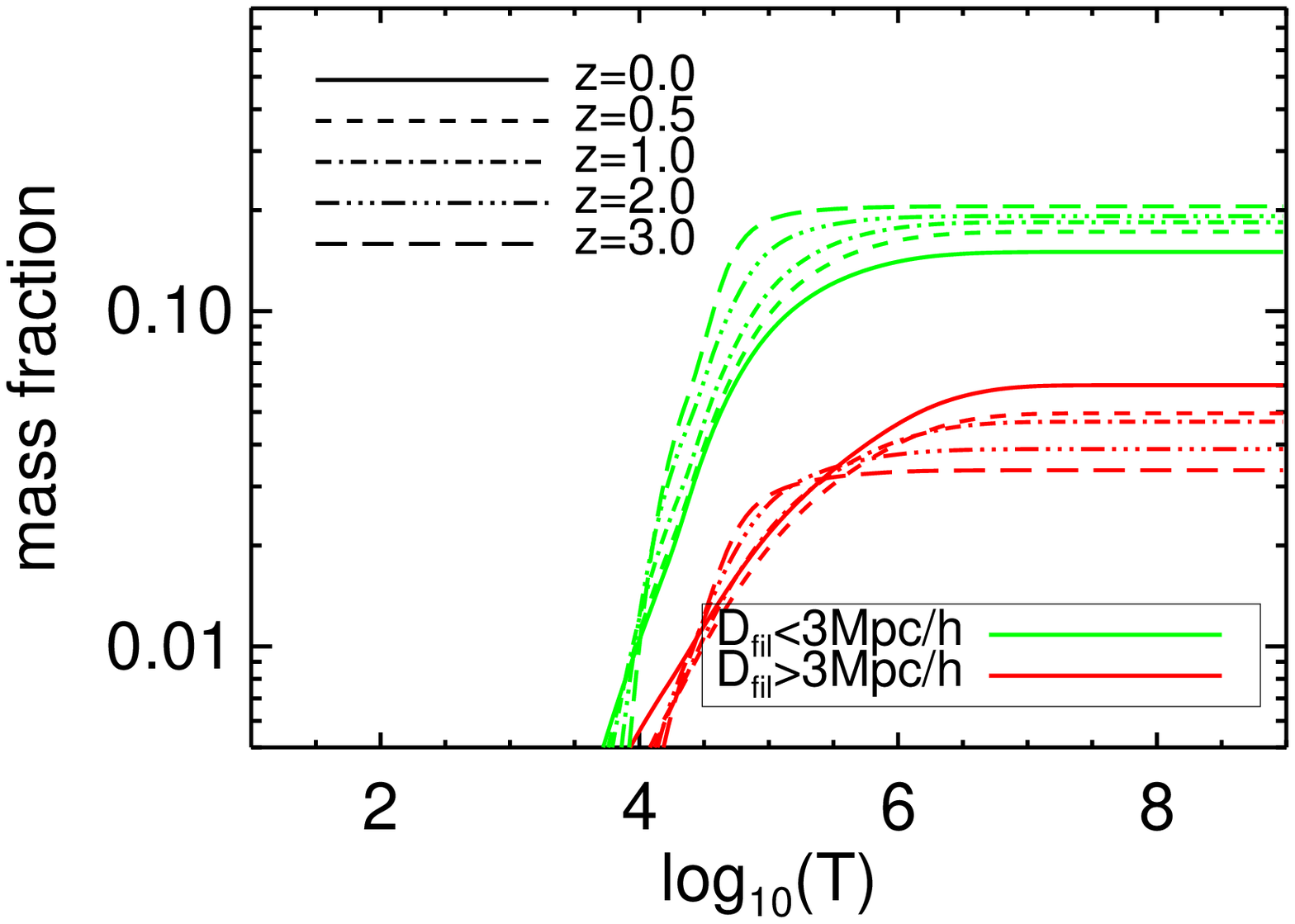}
\caption{The cumulative mass fraction of gas residing in filaments to all the intergalactic medium as a function of temperature. Green and red lines indicate filaments with width $\rm{D_{fil}<3\ Mpc}/h$ and $\rm{D_{fil}>3\ Mpc}/h$ respectively. Long dashed, three-dotted-dashed, dotted-dashed, dashed, and solid lines show results at $z=3.0, 2.0, 1.0, 0.5$ and $0.0$ respectively. Left: Gas within the virial radius $R_{200}$ of dark mass halos are excluded. Right: Gas within $2 \times R_{200}$ are exclueded.}
\label{fig:pdftem_tot}
\end{center}
\end{figure*}

Quantitative results can be found in the left panel of Figure \ref{fig:pdftem_tot}, which illustrates the cumulative mass fraction of the intergalactic medium residing in filaments as a function of temperature. Excluding the gases in halos hosted by filaments, the diffuse gases residing in filaments account for $\sim 34-38\%$ of the mass content of the IGM in the universe. The mass fraction of the IGM contributed by the gas in thick filaments with $\rm{D_{fil}>3\ Mpc}/h$ rises from $4.0\%$ at $z=3.0$ to $8.3\%$ at $z=1.0$, and further to $13.2\%$ at $z=0.0$. Meanwhile, the mass fraction contributed by gas in filaments with $\rm{D_{fil}<3\ Mpc}/h$ decreases gradually from $29.5\%$ at $z=3.0$ to $25.2\%$ at $z=0.0$. 

The mass fractions of gas hotter than $10^{5.5}$ and $10^{6.0}$ K are listed in Table \ref{tab:hot_frac}. $61.5\%$ of the gas in thick filaments are in the Warm and Hot phases, i.e., $T>10^{5.5}$ K at $z=0$, increased significantly from only $4.9\%$ at $z=3.0$. The counterpart fractions in thin filaments are much smaller, e.g. $23.8\%$ and $1.7\%$ at $z=0.0$ and $z=3.0$ respectively. At $z=3.0$, only $1.5\%$ of the gas in filaments with width $\rm{D_{fil}>3\ Mpc}/h$ are hotter than $10^6$ K. This fraction goes up to $7.7\%$, $25.6\%$, $38.0\%$ and $44.7\%$ at $z=2.0, 1.0, 0.5$ and $0.0$ respectively. The corresponding mass fractions of hot gas in filaments with width $\rm{D_{fil}<3\ Mpc}/h$ are $0.3\%$, $1.7\%$, $5.8\%$, $9.2\%$ and $11.2\%$ at redshift $3.0, 2.0, 1.0, 0.5$ and $0.0$ respectively. 

Note that, dark matter halos may also have nonnegligible influence on the thermal state of gas outside the virial radius. To further exclude such possible effect, we remove the gas within the sphere of radius $2R_{vir}$ sited at halo center and repeat the previous calculation. The corresponding cumulative mass fraction as a function of temperature is given in the right panel of Figure \ref{fig:pdftem_tot}. The bottom two rows in Table \ref{tab:hot_frac} summarize the mass fractions of gas that are hotter than $10^{5.5}$ and $10^{6}$ K in thick and thin filaments. We can see that, in regions relatively far from dark matter halos, thick filaments still host a much larger fraction of warm/hot gas than in thin filaments. This result agrees generally with previous studies. For instance, \citealt{2021A&A...649A.117G} claimed that the gas in relatively shorter and thicker filaments, found in more denser region, are hotter than gas in relatively longer and thinner filaments in the IllustrisTNG simulation. The average temperature in the core region of short filaments in \citealt{2021A&A...649A.117G} can be larger than $10^6$ K, if the gas within spheres of the radius $3\times R_{200}$ centred at massive halos is excluded. In contrast, the average temperatures for long filaments in less dense regions are around $3-4 \times 10^4$ K. 

\begin{deluxetable}{cccc}[htbp]
\tablenum{1}
\tablecaption{Mass fractions of gas that are hotter than $10^{5.5}\ $K, and $10^{6.0}\ $ K in thin and thick filaments. `$R_{vir}\ $excl.' and `$2R_{vir}\ $excl.' indicate gas within the sphere centered at the center of dark matter halos and of radius $R_{vir}$ and $2R_{vir}$ are excluded, respectively.}
\label{tab:hot_frac}
\tablewidth{0pt}
\tablehead{
\colhead{Gas location} &  \colhead{$z$} & \colhead{$\rm{T}>10^{5.5}$ K} &\colhead{$\rm{T}>10^{6.0}$ K} 
}
%\decimalcolnumbers
\startdata
{ }& 0.0 & 61.5\% & 44.7\% \\
thick filaments& 0.5 & 54.8\% & 38.0\% \\
($R_{vir}\ $ excl.)& 1.0 & 42.9\% & 25.6\%\\
{ }& 2.0 & 17.3\% & 7.7\% \\
{ }& 3.0 & 4.9\% & 1.5\% \\
\hline
{ }& 0.0 & 23.8\% & 11.2\% \\
thin filaments& 0.5 & 21.5\% & 9.2\% \\
($R_{vir}\ $excl.)& 1.0 & 16.1\% & 5.8\%\\
{ }& 2.0 & 6.3\% & 1.7\% \\
{ }& 3.0 & 1.7\% & 0.3\% \\
\hline
{ }& 0.0 & 45.1\% & 24.3\% \\
thick filaments & 0.5 & 39.7\% & 19.8\% \\
($2R_{vir}\ $excl.)& 1.0 & 30.8\% & 13.2\%\\
{ }& 2.0 & 12.6\% & 3.9\% \\
{ }& 3.0 & 4.9\% & 1.2\% \\
\hline
{ }& 0.0 & 20.3\% & 6.9\% \\
thin filaments& 0.5 & 19.6\% & 6.2\% \\
($2R_{vir}\ $excl.)& 1.0 & 15.2\% & 4.3\%\\
{ }& 2.0 & 5.7\% & 1.1\%\\
{ }& 3.0 & 2.0\% & 0.3\% \\
\enddata
%\tablecomments{.}
\end{deluxetable}

We argue that the evolution of gas phase in these two groups of filaments with different local widths could be at least partially responsible for the discrepancy on gas accretion rate onto hosted halos at $z \leq 0.5$. Halos in thick filaments with $\rm{D_{fil}>3\ Mpc}/h$, comprising one thirds of the halos locating in filaments, are surrounded by hotter ambient gases at $z \leq 0.5$, with respect to their counterparts in thin filaments. For halo with a given mass, it will be more difficult to capture hotter ambient gas around its gravitational well. Therefore, the gas accretion onto low mass halos in thick filaments would be suppressed, in comparison with those halos in thin filaments. This process is somewhat similar to the pre-heating scenario proposed in the literature (e.g \citealt{2002MNRAS.333..768M}; \citealt{2007MNRAS.377..617L}), yet it works only in the latest period with $z\leq 0.5$ versus $z \leq 2.0$ required in \cite{2002MNRAS.333..768M}, and only in a limited fraction of halos, i.e. $\sim 20\%$. The suppression of gas supply to halos in prominent filaments could provide a potential physical mechanism to understand the recent observation, which suggests that galaxies must have undergone pre-processing/processing in large scale filaments. For instance, \citealt{2021arXiv210104389C} found that the gas content is decreasing from field to filament and cluster for galaxies in and around the Virgo cluster.

In the end of the last subsection, we find that there is barely any difference on the gas accretion rate between halos in the inner and outer regions of thin filaments from early time to $z=0$. This feature holds for thick filaments till $z=0.5$. Yet, at $z=0$, for halos less massive than $10^{11.6}\ \rm{M_{\odot}}$ and residing in thick filaments, the gas accretion rate in the inner region is lower than that in the outer region by $\sim 50\%$. The probable reason is as follows. As has been shown in \citealt{2021ApJ...920....2Z}, for filaments with width $\rm{D_{fil}\lesssim 4.0\ Mpc}/h$, the gas temperature in filaments increases in a quite slow pace while moving from the outer region inward to the center along the direction that perpendicular to the spine. Therefore, for halos residing in the inner and outer regions of thin filaments, the temperature of ambient gas would be close, which would lead to minor difference on the gas accretion process.

Only in filaments with diameter greater than $4-5\ \rm{Mpc/h}$,  does the gas temperature grow significantly in the inner region. However, the number of filaments with $\rm{D_{fil}}$ greater than $4-5\ \rm{Mpc}/h$ is rather small till $z \sim 1.0$, and grows rapidly since then. It needs some time to accumulate enough hot gas in the inner region of thick filaments to have notable effect on the gas accretion process of halos. Consequently, the difference between gas accretion rate to halos in the inner and outer region of thick filaments remains subtle till $z=0.5$, and only become notable at $z=0.0$.

\section{Summary and Discussions} 
\label{sec:summary}
 In this work, we have probed the impact of comic filament on the gas accretion rate of dark matter halos residing in filaments based on a cosmological hydrodynamical simulation. We have measured the gas accretion rate onto halos massive than $10^{10.6}\  M_{\odot}$at $z\leq 4.0$. We focus on halos residing in filaments and split them into two groups according to two methods: one is depending on the distance between each halo's center to the spine of filaments, and the other is on the local diameter of filament where the halo is residing. We have investigated if there is any difference on the gas accretion rate between halos in different groups. We find that:
\begin{enumerate}
\item At $z \geq 1.0$, there is negligible difference on the gas accretion rate between halos residing in `thick' filaments with width $\rm{D_{fil}>3.0\ Mpc}/h$, and in `thin' filaments with $\rm{D_{fil}<3.0\ Mpc}/h$. Down to $z=0.5$, this difference became notable. At $z=0.0$, the gas accretion rate onto halos with $M_h<10^{12}\  M_{\odot}$ residing in thick filaments is suppressed significantly, by a factor of 2-3, with respect to their counterparts in thin filaments. Using the virial temperature of halos as the dividing threshold, both the gas accretion in cold and hot modes are subjected to this suppression.

\item From high redshifts down to $z=0.5$, the gas accretion rate onto halo depends very weakly on its perpendicular distance to the filament spine, i.e., whether the distance from the spine is smaller than or larger than $1\ \rm{Mpc}/h$. At $z=0.0$, the gas accretion onto halos far from the spine is mildly slower than those near the spline in the mass range of halos $\rm{M_h \lesssim 10^{12}\ M_{\odot}}$, which is due to halos far from the spine tend to residing in thick filaments in our samples.

\item{Halos in the inner and outer regions, measured by the rescaled distance to the spine, of thin filaments shows no difference on the gas accretion rate from high redshifts to $z=0$.  Similar feature happens in thick filaments till $z=0.5$. Later, for halos less massive than $10^{11.6}\ \rm{M_{\odot}}$, the gas accretion rate to halos in the inner region of thick filaments is lower than that in the outer region by $\sim 50\%$.}

\end{enumerate}

Moreover, we have explored the properties of gas in filaments and demonstrated that thick filaments are usually located in relatively high overdense regions, while thin filaments often appear in the transition areas between underdense voids and highly overdense regions. Since $z \sim 1.0$, a considerable fraction of the intergalactic gas residing in thick filaments with width $\rm{D_{fil}>3.0\ Mpc}/h$ has become hotter than $10^6$ K, heated by the gravitational collapse of prominent filaments. This fraction goes up rapidly as redshift decreasing, and reaches $45\%$ at $z=0$. In contrast, there is less gas in filaments with $\rm{D_{fil}<3.0\ Mpc}/h$ that have been heated up to $10^6$ K. This feature is broadly consistent with the thermal state of filaments gas in the IllustrisTNG simulation (\citealt{2021A&A...649A.117G}), where the average temperature of gas in the core region of shorter and thicker filaments is above $10^6$ K, and is larger than the average temperature in relatively longer and thinner filaments by a factor of 2-3. We argue that this discrepancy on the properties of ambient gas surrounding halos should be responsible for the difference on gas accretion rate onto halos at $z \lesssim 0.5 $ found in this work, at least partly if is not entirely. In the preheating model proposed by \cite{2002MNRAS.333..768M}, it has been suggested that the gas accretion rate onto halos residing in preheated ambient gas would be suppressed  (\citealt{2007MNRAS.377..617L}). 

However, our study shows that the gas accretion rate of halos in prominent filaments was suppressed only since $z\sim 0.5$. In addition, halos residing in thick filaments comprise one thirds of the halos hosting by filaments, i.e., $\sim 20\%$ of all the halos at $z<1$. Namely, the fraction of halos affected by hot ambient gas is limited, in comparison to the preheating model discussed in \cite{2002MNRAS.333..768M} and \cite{2007MNRAS.377..617L}. Nevertheless,  \cite{2017ApJ...837...16D} shows that the median star formation rate of galaxies in the COSMOS field gradually declines from field to cluster only after $z=0.8$, and this decline is not significant between field and filaments, but much evident between clusters and other regions. This trend is generally consistent with our finding on the impact of filaments on gas accretion rate. 

We argue that the suppression of gas accretion rate onto halos by the preheated gas in prominent cosmic filament at $z\leq 0.5$ could serve as a physical pre-processing/processing mechanism on a large scale to cut down the supply of gas to halos before they enter to massive groups and clusters . This large scale environment effect may further lower the gas reservoir in galaxies located in prominent cosmic filament.  Consequently it may be partially responsible for the observed transition of galaxies properties while approaching to cosmic filaments and nodes/clusters, such as the decreasing gas content and star formation activity, and becoming more passive/red (e.g. \citealt{2016MNRAS.457.2287A}; \citealt{2017MNRAS.466.1880C}; \citealt{2017ApJ...837...16D}; \citealt{2017A&A...600L...6K}; \citealt{2018MNRAS.474..547K};
\citealt{2019A&A...632A..49S};
\citealt{2020A&A...638A..75B}; \citealt{2021arXiv210513368W}; \citealt{2021arXiv210104389C}).

Yet, we expect that cosmic filaments with local diameter smaller than $\rm{3.0\ Mpc/h}$ would have minor effects on the halos and galaxies they host. To acquire a more robust result on the role of large scale cosmic filaments in shaping the properties of galaxies, it would be better to separate the cosmic filaments (or their segments) to subgroups according to their local width. Otherwise, the impact of prominent and tenuous filaments would be mixed and result in weak or null difference on the galaxies properties between the near-filament and control samples (e.g \citealt{2017MNRAS.466.4692K}). 

The cosmic web environment in this work is identified by the density field of spatial resolution $\sim 200\ h^{-1} \rm{kpc}$.  Therefore, our results is applicable for cosmic filaments with diameters larger than $ 200\ h^{-1} \rm{kpc}$, i.e generally larger than the size of dark matter halos of mass $\sim 10^{12} \rm{M_{\odot}}$. As \cite{2021ApJ...920....2Z} have shown that, more than $95\%$ of the gas in filaments are hosted by filaments thicker than $\rm{D_{fil}}=0.5\ Mpc/h$ in our samples. In addition, our work is mainly focusing on the influence of cosmic filaments at $z \leq 2$. For filaments with width comparable to, and smaller than a couple of $ 100\ h^{-1} \rm{kpc}$, i.e. the typical size of dark matter halos of mass $\sim 10^{12} \rm{M_{\odot}}$, their impacts on the gas accretion to halos have not been probed by our work. Actually, such filaments may enhance the accretion of gas to halos, especially for the gas accretion via cold mode at redshift higher than 2 (e.g. \citealt{2003MNRAS.345..349B}; \citealt{2005MNRAS.363....2K}; \citealt{2008MNRAS.390.1326O}; \citealt{2009Natur.457..451D}; \citealt{2013MNRAS.429.3353N}). Recently, based on a high-resolution zoom-in hydrodynamical simulation, \cite{2019MNRAS.485..464L} suggest that the filament with a diameter of tens of kpc and with shock temperature of a few times of $10^4$ K can assist gas cooling and enhance star formation in its residing dark matter halo at high redshifts $z=4.0$ and $z=2.5$. 

Finally, we should note that there are some limitations in our work. First of all, only one simulation is studied here. Further investigation on more simulation samples is needed to verify the results revealed in this work. Second, this study is focusing only on the gas accretion rate onto dark matter halos, and have not probed the gas accretion onto galaxies and its consequent effect on stellar component. However, for a robust investigation on galaxy scale, simulations are required to be implemented with more sophisticated sub-grid modules on star formation and feedback. Third, the method used to construct filaments cannot be applied to observed galaxy samples directly. In addition, a considerable fraction of the IGM residing in the filaments is not yet detected by observations. It needs much more efforts to probe the gas properties in filaments with different widths by observations, and justify the results presented in this work.  

\acknowledgments
We thank the anonymous referee for her/his useful comments
that improved the manuscript. This work is supported by the National Natural Science Foundation of China (NFSC) through grant 11733010. W.S.Z. is supported by NSFC grant 12173102, and by grant 202102080137 from Guangzhou Municipal Science and Technology Bureau. Z.F.P. is supported by 2021A1515012373 from Natural Science Foundation of Guangdong Province. The cosmological hydrodynamic simulation was performed on the Tianhe-II supercomputer. Post simulation analysis carried in this work was completed on the HPC facility of the School of Physics and Astronomy, Sun, Yat-Sen University.


\vspace{5mm}
%facilities{HST(STIS), Swift(XRT and UVOT), AAVSO, CTIO:1.3m,


%\software{RAMSES \citep{2002A&A...385..337T},  
%          Cloudy \citep{2013RMxAA..49..137F}, 
%          SExtractor \citep{1996A&AS..117..393B}
%          }



%\appendix


\bibliography{main}{}

\begin{thebibliography}{}
\expandafter\ifx\csname natexlab\endcsname\relax\def\natexlab#1{#1}\fi
\providecommand{\url}[1]{\href{#1}{#1}}
\providecommand{\dodoi}[1]{doi:~\href{http://doi.org/#1}{\nolinkurl{#1}}}
\providecommand{\doeprint}[1]{\href{http://ascl.net/#1}{\nolinkurl{http://ascl.net/#1}}}
\providecommand{\doarXiv}[1]{\href{https://arxiv.org/abs/#1}{\nolinkurl{https://arxiv.org/abs/#1}}}

\bibitem[{{Abadi} {et~al.}(1999){Abadi}, {Moore}, \&
  {Bower}}]{1999MNRAS.308..947A}
{Abadi}, M.~G., {Moore}, B., \& {Bower}, R.~G. 1999, \mnras, 308, 947,
  \dodoi{10.1046/j.1365-8711.1999.02715.x}

\bibitem[{{Alberts} {et~al.}(2016){Alberts}, {Pope}, {Brodwin}, {Chung},
  {Cybulski}, {Dey}, {Eisenhardt}, {Galametz}, {Gonzalez}, {Jannuzi},
  {Stanford}, {Snyder}, {Stern}, \& {Zeimann}}]{2016ApJ...825...72A}
{Alberts}, S., {Pope}, A., {Brodwin}, M., {et~al.} 2016, \apj, 825, 72,
  \dodoi{10.3847/0004-637X/825/1/72}

\bibitem[{{Alpaslan} {et~al.}(2014){Alpaslan}, {Robotham}, {Driver}, {Norberg},
  {Baldry}, {Bauer}, {Bland-Hawthorn}, {Brown}, {Cluver}, {Colless}, {Foster},
  {Hopkins}, {Van Kampen}, {Kelvin}, {Lara-Lopez}, {Liske}, {Lopez-Sanchez},
  {Loveday}, {McNaught-Roberts}, {Merson}, \& {Pimbblet}}]{2014MNRAS.438..177A}
{Alpaslan}, M., {Robotham}, A. S.~G., {Driver}, S., {et~al.} 2014, \mnras, 438,
  177, \dodoi{10.1093/mnras/stt2136}

\bibitem[{{Alpaslan} {et~al.}(2016){Alpaslan}, {Grootes}, {Marcum}, {Popescu},
  {Tuffs}, {Bland-Hawthorn}, {Brough}, {Brown}, {Davies}, {Driver}, {Holwerda},
  {Kelvin}, {Lara-L{\'o}pez}, {L{\'o}pez-S{\'a}nchez}, {Loveday}, {Moffett},
  {Taylor}, {Owers}, \& {Robotham}}]{2016MNRAS.457.2287A}
{Alpaslan}, M., {Grootes}, M., {Marcum}, P.~M., {et~al.} 2016, \mnras, 457,
  2287, \dodoi{10.1093/mnras/stw134}

\bibitem[{{Arag{\'o}n-Calvo} {et~al.}(2007){Arag{\'o}n-Calvo}, {Jones}, {van de
  Weygaert}, \& {van der Hulst}}]{2007A&A...474..315A}
{Arag{\'o}n-Calvo}, M.~A., {Jones}, B.~J.~T., {van de Weygaert}, R., \& {van
  der Hulst}, J.~M. 2007, \aap, 474, 315, \dodoi{10.1051/0004-6361:20077880}

\bibitem[{{Aragon Calvo} {et~al.}(2019){Aragon Calvo}, {Neyrinck}, \&
  {Silk}}]{2019OJAp....2E...7A}
{Aragon Calvo}, M.~A., {Neyrinck}, M.~C., \& {Silk}, J. 2019, The Open Journal
  of Astrophysics, 2, 7, \dodoi{10.21105/astro.1607.07881}

\bibitem[{{Arag{\'o}n-Calvo} {et~al.}(2010){Arag{\'o}n-Calvo}, {van de
  Weygaert}, \& {Jones}}]{2010MNRAS.408.2163A}
{Arag{\'o}n-Calvo}, M.~A., {van de Weygaert}, R., \& {Jones}, B. J.~T. 2010,
  \mnras, 408, 2163, \dodoi{10.1111/j.1365-2966.2010.17263.x}

\bibitem[{{Balogh} {et~al.}(2000){Balogh}, {Navarro}, \&
  {Morris}}]{2000ApJ...540..113B}
{Balogh}, M.~L., {Navarro}, J.~F., \& {Morris}, S.~L. 2000, \apj, 540, 113,
  \dodoi{10.1086/309323}

\bibitem[{{Barkana} \& {Loeb}(2001)}]{2001PhR...349..125B}
{Barkana}, R., \& {Loeb}, A. 2001, \physrep, 349, 125,
  \dodoi{10.1016/S0370-1573(01)00019-9}

\bibitem[{{Birnboim} \& {Dekel}(2003)}]{2003MNRAS.345..349B}
{Birnboim}, Y., \& {Dekel}, A. 2003, \mnras, 345, 349,
  \dodoi{10.1046/j.1365-8711.2003.06955.x}

\bibitem[{{Blanton} {et~al.}(2005){Blanton}, {Eisenstein}, {Hogg}, {Schlegel},
  \& {Brinkmann}}]{2005ApJ...629..143B}
{Blanton}, M.~R., {Eisenstein}, D., {Hogg}, D.~W., {Schlegel}, D.~J., \&
  {Brinkmann}, J. 2005, \apj, 629, 143, \dodoi{10.1086/422897}

\bibitem[{{Bond} {et~al.}(1996){Bond}, {Kofman}, \&
  {Pogosyan}}]{1996Natur.380..603B}
{Bond}, J.~R., {Kofman}, L., \& {Pogosyan}, D. 1996, \nat, 380, 603,
  \dodoi{10.1038/380603a0}

\bibitem[{{Bonjean} {et~al.}(2020){Bonjean}, {Aghanim}, {Douspis}, {Malavasi},
  \& {Tanimura}}]{2020A&A...638A..75B}
{Bonjean}, V., {Aghanim}, N., {Douspis}, M., {Malavasi}, N., \& {Tanimura}, H.
  2020, \aap, 638, A75, \dodoi{10.1051/0004-6361/201937313}

\bibitem[{{Calvi} {et~al.}(2013){Calvi}, {Poggianti}, {Vulcani}, \&
  {Fasano}}]{2013MNRAS.432.3141C}
{Calvi}, R., {Poggianti}, B.~M., {Vulcani}, B., \& {Fasano}, G. 2013, \mnras,
  432, 3141, \dodoi{10.1093/mnras/stt667}

\bibitem[{{Calvi} {et~al.}(2018){Calvi}, {Vulcani}, {Poggianti}, {Moretti},
  {Fritz}, \& {Fasano}}]{2018MNRAS.481.3456C}
{Calvi}, R., {Vulcani}, B., {Poggianti}, B.~M., {et~al.} 2018, \mnras, 481,
  3456, \dodoi{10.1093/mnras/sty2476}

\bibitem[{{Capak} {et~al.}(2007){Capak}, {Abraham}, {Ellis}, {Mobasher},
  {Scoville}, {Sheth}, \& {Koekemoer}}]{2007ApJS..172..284C}
{Capak}, P., {Abraham}, R.~G., {Ellis}, R.~S., {et~al.} 2007, \apjs, 172, 284,
  \dodoi{10.1086/518424}

\bibitem[{{Castignani} {et~al.}(2021){Castignani}, {Combes}, {Jablonka},
  {Finn}, {Rudnick}, {Vulcani}, {Desai}, {Zaritsky}, \&
  {Salom{\'e}}}]{2021arXiv210104389C}
{Castignani}, G., {Combes}, F., {Jablonka}, P., {et~al.} 2021, arXiv e-prints,
  arXiv:2101.04389.
\newblock \doarXiv{2101.04389}

\bibitem[{{Cautun} {et~al.}(2014){Cautun}, {van de Weygaert}, {Jones}, \&
  {Frenk}}]{2014MNRAS.441.2923C}
{Cautun}, M., {van de Weygaert}, R., {Jones}, B. J.~T., \& {Frenk}, C.~S. 2014,
  \mnras, 441, 2923, \dodoi{10.1093/mnras/stu768}

\bibitem[{{Chartab} {et~al.}(2020){Chartab}, {Mobasher}, {Darvish},
  {Finkelstein}, {Guo}, {Kodra}, {Lee}, {Newman}, {Pacifici}, {Papovich},
  {Sattari}, {Shahidi}, {Dickinson}, {Faber}, {Ferguson}, {Giavalisco}, \&
  {Jafariyazani}}]{2020ApJ...890....7C}
{Chartab}, N., {Mobasher}, B., {Darvish}, B., {et~al.} 2020, \apj, 890, 7,
  \dodoi{10.3847/1538-4357/ab61fd}

\bibitem[{{Chen} {et~al.}(2017){Chen}, {Ho}, {Mandelbaum}, {Bahcall},
  {Brownstein}, {Freeman}, {Genovese}, {Schneider}, \&
  {Wasserman}}]{2017MNRAS.466.1880C}
{Chen}, Y.-C., {Ho}, S., {Mandelbaum}, R., {et~al.} 2017, \mnras, 466, 1880,
  \dodoi{10.1093/mnras/stw3127}

\bibitem[{{Christlein} \& {Zabludoff}(2005)}]{2005ApJ...621..201C}
{Christlein}, D., \& {Zabludoff}, A.~I. 2005, \apj, 621, 201,
  \dodoi{10.1086/427427}

\bibitem[{{Colberg} {et~al.}(2005){Colberg}, {Krughoff}, \&
  {Connolly}}]{2005MNRAS.359..272C}
{Colberg}, J.~M., {Krughoff}, K.~S., \& {Connolly}, A.~J. 2005, \mnras, 359,
  272, \dodoi{10.1111/j.1365-2966.2005.08897.x}

\bibitem[{{Colless} {et~al.}(2003){Colless}, {Peterson}, {Jackson}, {Peacock},
  {Cole}, {Norberg}, {Baldry}, {Baugh}, {Bland-Hawthorn}, {Bridges}, {Cannon},
  {Collins}, {Couch}, {Cross}, {Dalton}, {De Propris}, {Driver}, {Efstathiou},
  {Ellis}, {Frenk}, {Glazebrook}, {Lahav}, {Lewis}, {Lumsden}, {Maddox},
  {Madgwick}, {Sutherland}, \& {Taylor}}]{2003astro.ph..6581C}
{Colless}, M., {Peterson}, B.~A., {Jackson}, C., {et~al.} 2003, arXiv e-prints,
  astro.
\newblock \doarXiv{astro-ph/0306581}

\bibitem[{{Cooper} {et~al.}(2008){Cooper}, {Newman}, {Weiner}, {Yan},
  {Willmer}, {Bundy}, {Coil}, {Conselice}, {Davis}, {Faber}, {Gerke},
  {Guhathakurta}, {Koo}, \& {Noeske}}]{2008MNRAS.383.1058C}
{Cooper}, M.~C., {Newman}, J.~A., {Weiner}, B.~J., {et~al.} 2008, \mnras, 383,
  1058, \dodoi{10.1111/j.1365-2966.2007.12613.x}

\bibitem[{{Cybulski} {et~al.}(2014){Cybulski}, {Yun}, {Fazio}, \&
  {Gutermuth}}]{2014MNRAS.439.3564C}
{Cybulski}, R., {Yun}, M.~S., {Fazio}, G.~G., \& {Gutermuth}, R.~A. 2014,
  \mnras, 439, 3564, \dodoi{10.1093/mnras/stu200}

\bibitem[{{Darvish} {et~al.}(2017){Darvish}, {Mobasher}, {Martin}, {Sobral},
  {Scoville}, {Stroe}, {Hemmati}, \& {Kartaltepe}}]{2017ApJ...837...16D}
{Darvish}, B., {Mobasher}, B., {Martin}, D.~C., {et~al.} 2017, \apj, 837, 16,
  \dodoi{10.3847/1538-4357/837/1/16}

\bibitem[{{Darvish} {et~al.}(2016){Darvish}, {Mobasher}, {Sobral}, {Rettura},
  {Scoville}, {Faisst}, \& {Capak}}]{2016ApJ...825..113D}
{Darvish}, B., {Mobasher}, B., {Sobral}, D., {et~al.} 2016, \apj, 825, 113,
  \dodoi{10.3847/0004-637X/825/2/113}

\bibitem[{{de Lapparent} {et~al.}(1986){de Lapparent}, {Geller}, \&
  {Huchra}}]{1986ApJ...302L...1D}
{de Lapparent}, V., {Geller}, M.~J., \& {Huchra}, J.~P. 1986, \apjl, 302, L1,
  \dodoi{10.1086/184625}

\bibitem[{{Dekel} {et~al.}(2009){Dekel}, {Birnboim}, {Engel}, {Freundlich},
  {Goerdt}, {Mumcuoglu}, {Neistein}, {Pichon}, {Teyssier}, \&
  {Zinger}}]{2009Natur.457..451D}
{Dekel}, A., {Birnboim}, Y., {Engel}, G., {et~al.} 2009, \nat, 457, 451,
  \dodoi{10.1038/nature07648}

\bibitem[{{Dressler}(1980)}]{1980ApJ...236..351D}
{Dressler}, A. 1980, \apj, 236, 351, \dodoi{10.1086/157753}

\bibitem[{{Dubois} {et~al.}(2014){Dubois}, {Pichon}, {Welker}, {Le Borgne},
  {Devriendt}, {Laigle}, {Codis}, {Pogosyan}, {Arnouts}, {Benabed}, {Bertin},
  {Blaizot}, {Bouchet}, {Cardoso}, {Colombi}, {de Lapparent}, {Desjacques},
  {Gavazzi}, {Kassin}, {Kimm}, {McCracken}, {Milliard}, {Peirani}, {Prunet},
  {Rouberol}, {Silk}, {Slyz}, {Sousbie}, {Teyssier}, {Tresse}, {Treyer},
  {Vibert}, \& {Volonteri}}]{2014MNRAS.444.1453D}
{Dubois}, Y., {Pichon}, C., {Welker}, C., {et~al.} 2014, \mnras, 444, 1453,
  \dodoi{10.1093/mnras/stu1227}

\bibitem[{{Elbaz} {et~al.}(2007){Elbaz}, {Daddi}, {Le Borgne}, {Dickinson},
  {Alexander}, {Chary}, {Starck}, {Brandt}, {Kitzbichler}, {MacDonald},
  {Nonino}, {Popesso}, {Stern}, \& {Vanzella}}]{2007A&A...468...33E}
{Elbaz}, D., {Daddi}, E., {Le Borgne}, D., {et~al.} 2007, \aap, 468, 33,
  \dodoi{10.1051/0004-6361:20077525}

\bibitem[{{Faucher-Gigu{\`e}re} {et~al.}(2011){Faucher-Gigu{\`e}re},
  {Kere{\v{s}}}, \& {Ma}}]{2011MNRAS.417.2982F}
{Faucher-Gigu{\`e}re}, C.-A., {Kere{\v{s}}}, D., \& {Ma}, C.-P. 2011, \mnras,
  417, 2982, \dodoi{10.1111/j.1365-2966.2011.19457.x}

\bibitem[{{Forero-Romero} {et~al.}(2009){Forero-Romero}, {Hoffman},
  {Gottl{\"o}ber}, {Klypin}, \& {Yepes}}]{2009MNRAS.396.1815F}
{Forero-Romero}, J.~E., {Hoffman}, Y., {Gottl{\"o}ber}, S., {Klypin}, A., \&
  {Yepes}, G. 2009, \mnras, 396, 1815, \dodoi{10.1111/j.1365-2966.2009.14885.x}

\bibitem[{{Gal{\'a}rraga-Espinosa} {et~al.}(2021){Gal{\'a}rraga-Espinosa},
  {Aghanim}, {Langer}, \& {Tanimura}}]{2021A&A...649A.117G}
{Gal{\'a}rraga-Espinosa}, D., {Aghanim}, N., {Langer}, M., \& {Tanimura}, H.
  2021, \aap, 649, A117, \dodoi{10.1051/0004-6361/202039781}

\bibitem[{{Gr{\"u}tzbauch} {et~al.}(2011{\natexlab{a}}){Gr{\"u}tzbauch},
  {Conselice}, {Varela}, {Bundy}, {Cooper}, {Skibba}, \&
  {Willmer}}]{2011MNRAS.411..929G}
{Gr{\"u}tzbauch}, R., {Conselice}, C.~J., {Varela}, J., {et~al.}
  2011{\natexlab{a}}, \mnras, 411, 929,
  \dodoi{10.1111/j.1365-2966.2010.17727.x}

\bibitem[{{Gr{\"u}tzbauch} {et~al.}(2011{\natexlab{b}}){Gr{\"u}tzbauch},
  {Conselice}, {Bauer}, {Bluck}, {Chuter}, {Buitrago}, {Mortlock}, {Weinzirl},
  \& {Jogee}}]{2011MNRAS.418..938G}
{Gr{\"u}tzbauch}, R., {Conselice}, C.~J., {Bauer}, A.~E., {et~al.}
  2011{\natexlab{b}}, \mnras, 418, 938,
  \dodoi{10.1111/j.1365-2966.2011.19559.x}

\bibitem[{{Gunn} \& {Gott}(1972)}]{1972ApJ...176....1G}
{Gunn}, J.~E., \& {Gott}, J.~Richard, I. 1972, \apj, 176, 1,
  \dodoi{10.1086/151605}

\bibitem[{{Guo} {et~al.}(2017){Guo}, {Bell}, {Lu}, {Koo}, {Faber}, {Koekemoer},
  {Kurczynski}, {Lee}, {Papovich}, {Chen}, {Dekel}, {Ferguson}, {Fontana},
  {Giavalisco}, {Kocevski}, {Nayyeri}, {P{\'e}rez-Gonz{\'a}lez}, {Pforr},
  {Rodr{\'\i}guez-Puebla}, \& {Santini}}]{2017ApJ...841L..22G}
{Guo}, Y., {Bell}, E.~F., {Lu}, Y., {et~al.} 2017, \apjl, 841, L22,
  \dodoi{10.3847/2041-8213/aa70e9}

\bibitem[{{Haardt} \& {Madau}(1996)}]{1996ApJ...461...20H}
{Haardt}, F., \& {Madau}, P. 1996, \apj, 461, 20, \dodoi{10.1086/177035}

\bibitem[{{Hahn} {et~al.}(2007){Hahn}, {Porciani}, {Carollo}, \&
  {Dekel}}]{2007MNRAS.375..489H}
{Hahn}, O., {Porciani}, C., {Carollo}, C.~M., \& {Dekel}, A. 2007, \mnras, 375,
  489, \dodoi{10.1111/j.1365-2966.2006.11318.x}

\bibitem[{{Haider} {et~al.}(2016){Haider}, {Steinhauser}, {Vogelsberger},
  {Genel}, {Springel}, {Torrey}, \& {Hernquist}}]{2016MNRAS.457.3024H}
{Haider}, M., {Steinhauser}, D., {Vogelsberger}, M., {et~al.} 2016, \mnras,
  457, 3024, \dodoi{10.1093/mnras/stw077}

\bibitem[{{Hellwing} {et~al.}(2021){Hellwing}, {Cautun}, {van de Weygaert}, \&
  {Jones}}]{2021PhRvD.103f3517H}
{Hellwing}, W.~A., {Cautun}, M., {van de Weygaert}, R., \& {Jones}, B.~T. 2021,
  \prd, 103, 063517, \dodoi{10.1103/PhysRevD.103.063517}

\bibitem[{{Hogg} {et~al.}(2003){Hogg}, {Blanton}, {Eisenstein}, {Gunn},
  {Schlegel}, {Zehavi}, {Bahcall}, {Brinkmann}, {Csabai}, {Schneider},
  {Weinberg}, \& {York}}]{2003ApJ...585L...5H}
{Hogg}, D.~W., {Blanton}, M.~R., {Eisenstein}, D.~J., {et~al.} 2003, \apjl,
  585, L5, \dodoi{10.1086/374238}

\bibitem[{{Icke}(1973)}]{1973A&A....27....1I}
{Icke}, V. 1973, \aap, 27, 1

\bibitem[{{Kauffmann} {et~al.}(2004){Kauffmann}, {White}, {Heckman},
  {M{\'e}nard}, {Brinchmann}, {Charlot}, {Tremonti}, \&
  {Brinkmann}}]{2004MNRAS.353..713K}
{Kauffmann}, G., {White}, S. D.~M., {Heckman}, T.~M., {et~al.} 2004, \mnras,
  353, 713, \dodoi{10.1111/j.1365-2966.2004.08117.x}

\bibitem[{{Kawinwanichakij} {et~al.}(2017){Kawinwanichakij}, {Papovich},
  {Quadri}, {Glazebrook}, {Kacprzak}, {Allen}, {Bell}, {Croton}, {Dekel},
  {Ferguson}, {Forrest}, {Grogin}, {Guo}, {Kocevski}, {Koekemoer}, {Labb{\'e}},
  {Lucas}, {Nanayakkara}, {Spitler}, {Straatman}, {Tran}, {Tomczak}, \& {van
  Dokkum}}]{2017ApJ...847..134K}
{Kawinwanichakij}, L., {Papovich}, C., {Quadri}, R.~F., {et~al.} 2017, \apj,
  847, 134, \dodoi{10.3847/1538-4357/aa8b75}

\bibitem[{{Kere{\v{s}}} {et~al.}(2005){Kere{\v{s}}}, {Katz}, {Weinberg}, \&
  {Dav{\'e}}}]{2005MNRAS.363....2K}
{Kere{\v{s}}}, D., {Katz}, N., {Weinberg}, D.~H., \& {Dav{\'e}}, R. 2005,
  \mnras, 363, 2, \dodoi{10.1111/j.1365-2966.2005.09451.x}

\bibitem[{{Kleiner} {et~al.}(2017){Kleiner}, {Pimbblet}, {Jones}, {Koribalski},
  \& {Serra}}]{2017MNRAS.466.4692K}
{Kleiner}, D., {Pimbblet}, K.~A., {Jones}, D.~H., {Koribalski}, B.~S., \&
  {Serra}, P. 2017, \mnras, 466, 4692, \dodoi{10.1093/mnras/stw3328}

\bibitem[{{Kraljic} {et~al.}(2018){Kraljic}, {Arnouts}, {Pichon}, {Laigle}, {de
  la Torre}, {Vibert}, {Cadiou}, {Dubois}, {Treyer}, {Schimd}, {Codis}, {de
  Lapparent}, {Devriendt}, {Hwang}, {Le Borgne}, {Malavasi}, {Milliard},
  {Musso}, {Pogosyan}, {Alpaslan}, {Bland-Hawthorn}, \&
  {Wright}}]{2018MNRAS.474..547K}
{Kraljic}, K., {Arnouts}, S., {Pichon}, C., {et~al.} 2018, \mnras, 474, 547,
  \dodoi{10.1093/mnras/stx2638}

\bibitem[{{Kuutma} {et~al.}(2017){Kuutma}, {Tamm}, \&
  {Tempel}}]{2017A&A...600L...6K}
{Kuutma}, T., {Tamm}, A., \& {Tempel}, E. 2017, \aap, 600, L6,
  \dodoi{10.1051/0004-6361/201730526}

\bibitem[{{Larson} {et~al.}(1980){Larson}, {Tinsley}, \&
  {Caldwell}}]{1980ApJ...237..692L}
{Larson}, R.~B., {Tinsley}, B.~M., \& {Caldwell}, C.~N. 1980, \apj, 237, 692,
  \dodoi{10.1086/157917}

\bibitem[{{Lee} {et~al.}(2021){Lee}, {Kim}, {Rey}, \&
  {Chung}}]{2021ApJ...906...68L}
{Lee}, Y., {Kim}, S., {Rey}, S.-C., \& {Chung}, J. 2021, \apj, 906, 68,
  \dodoi{10.3847/1538-4357/abcaa0}

\bibitem[{{Liao} \& {Gao}(2019)}]{2019MNRAS.485..464L}
{Liao}, S., \& {Gao}, L. 2019, \mnras, 485, 464, \dodoi{10.1093/mnras/stz441}

\bibitem[{{Lu} \& {Mo}(2007)}]{2007MNRAS.377..617L}
{Lu}, Y., \& {Mo}, H.~J. 2007, \mnras, 377, 617,
  \dodoi{10.1111/j.1365-2966.2007.11627.x}

\bibitem[{{Mahajan} {et~al.}(2018){Mahajan}, {Singh}, \&
  {Shobhana}}]{2018MNRAS.478.4336M}
{Mahajan}, S., {Singh}, A., \& {Shobhana}, D. 2018, \mnras, 478, 4336,
  \dodoi{10.1093/mnras/sty1370}

\bibitem[{{Martizzi} {et~al.}(2019){Martizzi}, {Vogelsberger}, {Artale},
  {Haider}, {Torrey}, {Marinacci}, {Nelson}, {Pillepich}, {Weinberger},
  {Hernquist}, {Naiman}, \& {Springel}}]{2019MNRAS.486.3766M}
{Martizzi}, D., {Vogelsberger}, M., {Artale}, M.~C., {et~al.} 2019, \mnras,
  486, 3766, \dodoi{10.1093/mnras/stz1106}

\bibitem[{{Mo} \& {Mao}(2002)}]{2002MNRAS.333..768M}
{Mo}, H.~J., \& {Mao}, S. 2002, \mnras, 333, 768,
  \dodoi{10.1046/j.1365-8711.2002.05416.x}

\bibitem[{{Moore} {et~al.}(1996){Moore}, {Katz}, {Lake}, {Dressler}, \&
  {Oemler}}]{1996Natur.379..613M}
{Moore}, B., {Katz}, N., {Lake}, G., {Dressler}, A., \& {Oemler}, A. 1996,
  \nat, 379, 613, \dodoi{10.1038/379613a0}

\bibitem[{{Nelson} {et~al.}(2013){Nelson}, {Vogelsberger}, {Genel}, {Sijacki},
  {Kere{\v{s}}}, {Springel}, \& {Hernquist}}]{2013MNRAS.429.3353N}
{Nelson}, D., {Vogelsberger}, M., {Genel}, S., {et~al.} 2013, \mnras, 429,
  3353, \dodoi{10.1093/mnras/sts595}

\bibitem[{{Ocvirk} {et~al.}(2008){Ocvirk}, {Pichon}, \&
  {Teyssier}}]{2008MNRAS.390.1326O}
{Ocvirk}, P., {Pichon}, C., \& {Teyssier}, R. 2008, \mnras, 390, 1326,
  \dodoi{10.1111/j.1365-2966.2008.13763.x}

\bibitem[{{Old} {et~al.}(2020){Old}, {Balogh}, {van der Burg}, {Biviano},
  {Yee}, {Pintos-Castro}, {Webb}, {Muzzin}, {Rudnick}, {Vulcani}, {Poggianti},
  {Cooper}, {Zaritsky}, {Cerulo}, {Wilson}, {Chan}, {Lidman}, {McGee},
  {Demarco}, {Forrest}, {De Lucia}, {Gilbank}, {Kukstas}, {McCarthy},
  {Jablonka}, {Nantais}, {Noble}, {Reeves}, \& {Shipley}}]{2020MNRAS.493.5987O}
{Old}, L.~J., {Balogh}, M.~L., {van der Burg}, R. F.~J., {et~al.} 2020, \mnras,
  493, 5987, \dodoi{10.1093/mnras/staa579}

\bibitem[{{Patel} {et~al.}(2009){Patel}, {Holden}, {Kelson}, {Illingworth}, \&
  {Franx}}]{2009ApJ...705L..67P}
{Patel}, S.~G., {Holden}, B.~P., {Kelson}, D.~D., {Illingworth}, G.~D., \&
  {Franx}, M. 2009, \apjl, 705, L67, \dodoi{10.1088/0004-637X/705/1/L67}

\bibitem[{{Peng} {et~al.}(2015){Peng}, {Maiolino}, \&
  {Cochrane}}]{2015Natur.521..192P}
{Peng}, Y., {Maiolino}, R., \& {Cochrane}, R. 2015, \nat, 521, 192,
  \dodoi{10.1038/nature14439}

\bibitem[{{Peng} {et~al.}(2010){Peng}, {Lilly}, {Kova{\v{c}}}, {Bolzonella},
  {Pozzetti}, {Renzini}, {Zamorani}, {Ilbert}, {Knobel}, {Iovino}, {Maier},
  {Cucciati}, {Tasca}, {Carollo}, {Silverman}, {Kampczyk}, {de Ravel},
  {Sanders}, {Scoville}, {Contini}, {Mainieri}, {Scodeggio}, {Kneib}, {Le
  F{\`e}vre}, {Bardelli}, {Bongiorno}, {Caputi}, {Coppa}, {de la Torre},
  {Franzetti}, {Garilli}, {Lamareille}, {Le Borgne}, {Le Brun}, {Mignoli},
  {Perez Montero}, {Pello}, {Ricciardelli}, {Tanaka}, {Tresse}, {Vergani},
  {Welikala}, {Zucca}, {Oesch}, {Abbas}, {Barnes}, {Bordoloi}, {Bottini},
  {Cappi}, {Cassata}, {Cimatti}, {Fumana}, {Hasinger}, {Koekemoer},
  {Leauthaud}, {Maccagni}, {Marinoni}, {McCracken}, {Memeo}, {Meneux}, {Nair},
  {Porciani}, {Presotto}, \& {Scaramella}}]{2010ApJ...721..193P}
{Peng}, Y.-j., {Lilly}, S.~J., {Kova{\v{c}}}, K., {et~al.} 2010, \apj, 721,
  193, \dodoi{10.1088/0004-637X/721/1/193}

\bibitem[{{Planck Collaboration} {et~al.}(2014){Planck Collaboration}, {Ade},
  {Aghanim}, {Armitage-Caplan}, {Arnaud}, {Ashdown}, {Atrio-Barand ela},
  {Aumont}, {Baccigalupi}, {Banday}, {Barreiro}, {Bartlett}, {Battaner},
  {Benabed}, {Beno{\^\i}t}, {Benoit-L{\'e}vy}, {Bernard}, {Bersanelli},
  {Bielewicz}, {Bobin}, {Bock}, {Bonaldi}, {Bond}, {Borrill}, {Bouchet},
  {Bridges}, {Bucher}, {Burigana}, {Butler}, {Calabrese}, {Cappellini},
  {Cardoso}, {Catalano}, {Challinor}, {Chamballu}, {Chary}, {Chen}, {Chiang},
  {Chiang}, {Christensen}, {Church}, {Clements}, {Colombi}, {Colombo},
  {Couchot}, {Coulais}, {Crill}, {Curto}, {Cuttaia}, {Danese}, {Davies},
  {Davis}, {de Bernardis}, {de Rosa}, {de Zotti}, {Delabrouille}, {Delouis},
  {D{\'e}sert}, {Dickinson}, {Diego}, {Dolag}, {Dole}, {Donzelli}, {Dor{\'e}},
  {Douspis}, {Dunkley}, {Dupac}, {Efstathiou}, {Elsner}, {En{\ss}lin},
  {Eriksen}, {Finelli}, {Forni}, {Frailis}, {Fraisse}, {Franceschi}, {Gaier},
  {Galeotta}, {Galli}, {Ganga}, {Giard}, {Giardino}, {Giraud-H{\'e}raud},
  {Gjerl{\o}w}, {Gonz{\'a}lez-Nuevo}, {G{\'o}rski}, {Gratton}, {Gregorio},
  {Gruppuso}, {Gudmundsson}, {Haissinski}, {Hamann}, {Hansen}, {Hanson},
  {Harrison}, {Henrot-Versill{\'e}}, {Hern{\'a}ndez-Monteagudo}, {Herranz},
  {Hildebrand t}, {Hivon}, {Hobson}, {Holmes}, {Hornstrup}, {Hou}, {Hovest},
  {Huffenberger}, {Jaffe}, {Jaffe}, {Jewell}, {Jones}, {Juvela},
  {Keih{\"a}nen}, {Keskitalo}, {Kisner}, {Kneissl}, {Knoche}, {Knox}, {Kunz},
  {Kurki-Suonio}, {Lagache}, {L{\"a}hteenm{\"a}ki}, {Lamarre}, {Lasenby},
  {Lattanzi}, {Laureijs}, {Lawrence}, {Leach}, {Leahy}, {Leonardi},
  {Le{\'o}n-Tavares}, {Lesgourgues}, {Lewis}, {Liguori}, {Lilje},
  {Linden-V{\o}rnle}, {L{\'o}pez-Caniego}, {Lubin}, {Mac{\'\i}as-P{\'e}rez},
  {Maffei}, {Maino}, {Mand olesi}, {Maris}, {Marshall}, {Martin},
  {Mart{\'\i}nez-Gonz{\'a}lez}, {Masi}, {Massardi}, {Matarrese}, {Matthai},
  {Mazzotta}, {Meinhold}, {Melchiorri}, {Melin}, {Mendes}, {Menegoni},
  {Mennella}, {Migliaccio}, {Millea}, {Mitra}, {Miville-Desch{\^e}nes},
  {Moneti}, {Montier}, {Morgante}, {Mortlock}, {Moss}, {Munshi}, {Murphy},
  {Naselsky}, {Nati}, {Natoli}, {Netterfield}, {N{\o}rgaard-Nielsen},
  {Noviello}, {Novikov}, {Novikov}, {O'Dwyer}, {Osborne}, {Oxborrow}, {Paci},
  {Pagano}, {Pajot}, {Paladini}, {Paoletti}, {Partridge}, {Pasian},
  {Patanchon}, {Pearson}, {Pearson}, {Peiris}, {Perdereau}, {Perotto},
  {Perrotta}, {Pettorino}, {Piacentini}, {Piat}, {Pierpaoli}, {Pietrobon},
  {Plaszczynski}, {Platania}, {Pointecouteau}, {Polenta}, {Ponthieu}, {Popa},
  {Poutanen}, {Pratt}, {Pr{\'e}zeau}, {Prunet}, {Puget}, {Rachen}, {Reach},
  {Rebolo}, {Reinecke}, {Remazeilles}, {Renault}, {Ricciardi}, {Riller},
  {Ristorcelli}, {Rocha}, {Rosset}, {Roudier}, {Rowan-Robinson},
  {Rubi{\~n}o-Mart{\'\i}n}, {Rusholme}, {Sandri}, {Santos}, {Savelainen},
  {Savini}, {Scott}, {Seiffert}, {Shellard}, {Spencer}, {Starck}, {Stolyarov},
  {Stompor}, {Sudiwala}, {Sunyaev}, {Sureau}, {Sutton}, {Suur-Uski}, {Sygnet},
  {Tauber}, {Tavagnacco}, {Terenzi}, {Toffolatti}, {Tomasi}, {Tristram},
  {Tucci}, {Tuovinen}, {T{\"u}rler}, {Umana}, {Valenziano}, {Valiviita}, {Van
  Tent}, {Vielva}, {Villa}, {Vittorio}, {Wade}, {Wandelt}, {Wehus}, {White},
  {White}, {Wilkinson}, {Yvon}, {Zacchei}, \& {Zonca}}]{2014A&A...571A..16P}
{Planck Collaboration}, {Ade}, P.~A.~R., {Aghanim}, N., {et~al.} 2014, \aap,
  571, A16, \dodoi{10.1051/0004-6361/201321591}

\bibitem[{{Poggianti} {et~al.}(2008){Poggianti}, {Desai}, {Finn}, {Bamford},
  {De Lucia}, {Varela}, {Arag{\'o}n-Salamanca}, {Halliday}, {Noll}, {Saglia},
  {Zaritsky}, {Best}, {Clowe}, {Milvang-Jensen}, {Jablonka}, {Pell{\'o}},
  {Rudnick}, {Simard}, {von der Linden}, \& {White}}]{2008ApJ...684..888P}
{Poggianti}, B.~M., {Desai}, V., {Finn}, R., {et~al.} 2008, \apj, 684, 888,
  \dodoi{10.1086/589936}

\bibitem[{{Quilis} {et~al.}(2000){Quilis}, {Moore}, \&
  {Bower}}]{2000Sci...288.1617Q}
{Quilis}, V., {Moore}, B., \& {Bower}, R. 2000, Science, 288, 1617,
  \dodoi{10.1126/science.288.5471.1617}

\bibitem[{{Rost} {et~al.}(2021){Rost}, {Kuchner}, {Welker}, {Pearce},
  {Stasyszyn}, {Gray}, {Cui}, {Dave}, {Knebe}, {Yepes}, \&
  {Rasia}}]{2021MNRAS.502..714R}
{Rost}, A., {Kuchner}, U., {Welker}, C., {et~al.} 2021, \mnras, 502, 714,
  \dodoi{10.1093/mnras/staa3792}

\bibitem[{{Sarron} {et~al.}(2019){Sarron}, {Adami}, {Durret}, \&
  {Laigle}}]{2019A&A...632A..49S}
{Sarron}, F., {Adami}, C., {Durret}, F., \& {Laigle}, C. 2019, \aap, 632, A49,
  \dodoi{10.1051/0004-6361/201935394}

\bibitem[{{Schaye} {et~al.}(2015){Schaye}, {Crain}, {Bower}, {Furlong},
  {Schaller}, {Theuns}, {Dalla Vecchia}, {Frenk}, {McCarthy}, {Helly},
  {Jenkins}, {Rosas-Guevara}, {White}, {Baes}, {Booth}, {Camps}, {Navarro},
  {Qu}, {Rahmati}, {Sawala}, {Thomas}, \& {Trayford}}]{2015MNRAS.446..521S}
{Schaye}, J., {Crain}, R.~A., {Bower}, R.~G., {et~al.} 2015, \mnras, 446, 521,
  \dodoi{10.1093/mnras/stu2058}

\bibitem[{{Seth} \& {Raychaudhury}(2020)}]{2020MNRAS.497..466S}
{Seth}, R., \& {Raychaudhury}, S. 2020, \mnras, 497, 466,
  \dodoi{10.1093/mnras/staa1779}

\bibitem[{{Simha} {et~al.}(2009){Simha}, {Weinberg}, {Dav{\'e}}, {Gnedin},
  {Katz}, \& {Kere{\v{s}}}}]{2009MNRAS.399..650S}
{Simha}, V., {Weinberg}, D.~H., {Dav{\'e}}, R., {et~al.} 2009, \mnras, 399,
  650, \dodoi{10.1111/j.1365-2966.2009.15341.x}

\bibitem[{{Singh} {et~al.}(2019){Singh}, {Gulati}, \&
  {Bagla}}]{2019MNRAS.489.5582S}
{Singh}, A., {Gulati}, M., \& {Bagla}, J.~S. 2019, \mnras, 489, 5582,
  \dodoi{10.1093/mnras/stz2523}

\bibitem[{{Singh} {et~al.}(2020){Singh}, {Mahajan}, \&
  {Bagla}}]{2020MNRAS.497.2265S}
{Singh}, A., {Mahajan}, S., \& {Bagla}, J.~S. 2020, \mnras, 497, 2265,
  \dodoi{10.1093/mnras/staa1913}

\bibitem[{{Skibba} {et~al.}(2009){Skibba}, {Bamford}, {Nichol}, {Lintott},
  {Andreescu}, {Edmondson}, {Murray}, {Raddick}, {Schawinski}, {Slosar},
  {Szalay}, {Thomas}, \& {Vandenberg}}]{2009MNRAS.399..966S}
{Skibba}, R.~A., {Bamford}, S.~P., {Nichol}, R.~C., {et~al.} 2009, \mnras, 399,
  966, \dodoi{10.1111/j.1365-2966.2009.15334.x}

\bibitem[{{Song} {et~al.}(2021){Song}, {Laigle}, {Hwang}, {Devriendt},
  {Dubois}, {Kraljic}, {Pichon}, {Slyz}, \& {Smith}}]{2021MNRAS.501.4635S}
{Song}, H., {Laigle}, C., {Hwang}, H.~S., {et~al.} 2021, \mnras, 501, 4635,
  \dodoi{10.1093/mnras/staa3981}

\bibitem[{{Springel} {et~al.}(2005){Springel}, {White}, {Jenkins}, {Frenk},
  {Yoshida}, {Gao}, {Navarro}, {Thacker}, {Croton}, {Helly}, {Peacock}, {Cole},
  {Thomas}, {Couchman}, {Evrard}, {Colberg}, \& {Pearce}}]{2005Natur.435..629S}
{Springel}, V., {White}, S. D.~M., {Jenkins}, A., {et~al.} 2005, \nat, 435,
  629, \dodoi{10.1038/nature03597}

\bibitem[{{Steyrleithner} {et~al.}(2020){Steyrleithner}, {Hensler}, \&
  {Boselli}}]{2020MNRAS.494.1114S}
{Steyrleithner}, P., {Hensler}, G., \& {Boselli}, A. 2020, \mnras, 494, 1114,
  \dodoi{10.1093/mnras/staa775}

\bibitem[{{Tegmark} {et~al.}(2004){Tegmark}, {Blanton}, {Strauss}, {Hoyle},
  {Schlegel}, {Scoccimarro}, {Vogeley}, {Weinberg}, {Zehavi}, {Berlind},
  {Budavari}, {Connolly}, {Eisenstein}, {Finkbeiner}, {Frieman}, {Gunn},
  {Hamilton}, {Hui}, {Jain}, {Johnston}, {Kent}, {Lin}, {Nakajima}, {Nichol},
  {Ostriker}, {Pope}, {Scranton}, {Seljak}, {Sheth}, {Stebbins}, {Szalay},
  {Szapudi}, {Verde}, {Xu}, {Annis}, {Bahcall}, {Brinkmann}, {Burles},
  {Castander}, {Csabai}, {Loveday}, {Doi}, {Fukugita}, {Gott}, {Hennessy},
  {Hogg}, {Ivezi{\'c}}, {Knapp}, {Lamb}, {Lee}, {Lupton}, {McKay}, {Kunszt},
  {Munn}, {O'Connell}, {Peoples}, {Pier}, {Richmond}, {Rockosi}, {Schneider},
  {Stoughton}, {Tucker}, {Vanden Berk}, {Yanny}, {York}, \& {SDSS
  Collaboration}}]{2004ApJ...606..702T}
{Tegmark}, M., {Blanton}, M.~R., {Strauss}, M.~A., {et~al.} 2004, \apj, 606,
  702, \dodoi{10.1086/382125}

\bibitem[{{Teyssier}(2002)}]{2002A&A...385..337T}
{Teyssier}, R. 2002, \aap, 385, 337, \dodoi{10.1051/0004-6361:20011817}

\bibitem[{{Tonnesen} \& {Bryan}(2009)}]{2009ApJ...694..789T}
{Tonnesen}, S., \& {Bryan}, G.~L. 2009, \apj, 694, 789,
  \dodoi{10.1088/0004-637X/694/2/789}

\bibitem[{{van de Voort} {et~al.}(2017){van de Voort}, {Bah{\'e}}, {Bower},
  {Correa}, {Crain}, {Schaye}, \& {Theuns}}]{2017MNRAS.466.3460V}
{van de Voort}, F., {Bah{\'e}}, Y.~M., {Bower}, R.~G., {et~al.} 2017, \mnras,
  466, 3460, \dodoi{10.1093/mnras/stw3356}

\bibitem[{{van de Voort} {et~al.}(2011){van de Voort}, {Schaye}, {Booth},
  {Haas}, \& {Dalla Vecchia}}]{2011MNRAS.414.2458V}
{van de Voort}, F., {Schaye}, J., {Booth}, C.~M., {Haas}, M.~R., \& {Dalla
  Vecchia}, C. 2011, \mnras, 414, 2458,
  \dodoi{10.1111/j.1365-2966.2011.18565.x}

\bibitem[{{van de Weygaert} \& {Bond}(2008)}]{2008LNP...740..335V}
{van de Weygaert}, R., \& {Bond}, J.~R. 2008, {Clusters and the Theory of the
  Cosmic Web}, ed. M.~{Plionis}, O.~{L{\'o}pez-Cruz}, \& D.~{Hughes}, Vol. 740,
  335, \dodoi{10.1007/978-1-4020-6941-3\_10}

\bibitem[{{Vogelsberger} {et~al.}(2014){Vogelsberger}, {Genel}, {Springel},
  {Torrey}, {Sijacki}, {Xu}, {Snyder}, {Nelson}, \&
  {Hernquist}}]{2014MNRAS.444.1518V}
{Vogelsberger}, M., {Genel}, S., {Springel}, V., {et~al.} 2014, \mnras, 444,
  1518, \dodoi{10.1093/mnras/stu1536}

\bibitem[{{Vulcani} {et~al.}(2012){Vulcani}, {Poggianti}, {Fasano}, {Desai},
  {Dressler}, {Oemler}, {Calvi}, {D'Onofrio}, \&
  {Moretti}}]{2012MNRAS.420.1481V}
{Vulcani}, B., {Poggianti}, B.~M., {Fasano}, G., {et~al.} 2012, \mnras, 420,
  1481, \dodoi{10.1111/j.1365-2966.2011.20135.x}

\bibitem[{{Vulcani} {et~al.}(2013){Vulcani}, {Poggianti}, {Oemler}, {Dressler},
  {Arag{\'o}n-Salamanca}, {De Lucia}, {Moretti}, {Gladders}, {Abramson}, \&
  {Halliday}}]{2013A&A...550A..58V}
{Vulcani}, B., {Poggianti}, B.~M., {Oemler}, A., {et~al.} 2013, \aap, 550, A58,
  \dodoi{10.1051/0004-6361/201118388}

\bibitem[{{White} \& {Rees}(1978)}]{1978MNRAS.183..341W}
{White}, S.~D.~M., \& {Rees}, M.~J. 1978, \mnras, 183, 341,
  \dodoi{10.1093/mnras/183.3.341}

\bibitem[{{White} \& {Silk}(1979)}]{1979ApJ...231....1W}
{White}, S.~D.~M., \& {Silk}, J. 1979, \apj, 231, 1, \dodoi{10.1086/157156}

\bibitem[{{Winkel} {et~al.}(2021){Winkel}, {Pasquali}, {Kraljic}, {Smith},
  {Gallazzi}, \& {Jackson}}]{2021arXiv210513368W}
{Winkel}, N., {Pasquali}, A., {Kraljic}, K., {et~al.} 2021, arXiv e-prints,
  arXiv:2105.13368.
\newblock \doarXiv{2105.13368}

\bibitem[{{Xu} {et~al.}(2020){Xu}, {Guo}, {Zheng}, {Gao}, {Lacey}, {Gu},
  {Liao}, {Shao}, {Mao}, {Zhang}, \& {Chen}}]{2020MNRAS.498.1839X}
{Xu}, W., {Guo}, Q., {Zheng}, H., {et~al.} 2020, \mnras, 498, 1839,
  \dodoi{10.1093/mnras/staa2497}

\bibitem[{{Zel'Dovich}(1970)}]{1970A&A.....5...84Z}
{Zel'Dovich}, Y.~B. 1970, \aap, 500, 13

\bibitem[{{Zhu} \& {Feng}(2017)}]{2017ApJ...838...21Z}
{Zhu}, W., \& {Feng}, L.-L. 2017, \apj, 838, 21,
  \dodoi{10.3847/1538-4357/aa61f9}

\bibitem[{{Zhu} \& {Feng}(2021)}]{2021ApJ...906...95Z}
---. 2021, \apj, 906, 95, \dodoi{10.3847/1538-4357/abcb90}

\bibitem[{{Zhu} {et~al.}(2021){Zhu}, {Zhang}, \& {Feng}}]{2021ApJ...920....2Z}
{Zhu}, W., {Zhang}, F., \& {Feng}, L.-L. 2021, \apj, 920, 2,
  \dodoi{10.3847/1538-4357/ac15f1}

\end{thebibliography}
\bibliographystyle{aasjournal}



\end{document}